\documentclass[conference]{IEEEtran}
\usepackage{xspace}
\usepackage{amsmath}
\usepackage{amsfonts}
\usepackage{amssymb}
\usepackage{epsfig}
\usepackage{algorithm}
\usepackage{algorithmic}

\def\ps@headings{%
\def\@oddhead{\mbox{}\scriptsize\rightmark \hfil \thepage}%
\def\@evenhead{\scriptsize\thepage \hfil \leftmark\mbox{}}%
\def\@oddfoot{}%
\def\@evenfoot{}}
\makeatother
\usepackage{subfigure}
                                                                                                                                                             
\newcommand{\ignore}[1]{}

\usepackage{url}
\usepackage{subfigure}

\newcommand{\threshold}{\ensuremath{\theta}\xspace}

\newcommand{\funcs}[2]{\ensuremath{\mathbf{Func}({#1} \rightarrow {#2})}\xspace}

\newcommand{\prf}[1]{\ensuremath{f_{#1}}\xspace}
\newcommand{\key}{\ensuremath{K}\xspace}

\newcommand{\prfone}[1]{\ensuremath{\prf{#1}^1}\xspace}
\newcommand{\keyone}{\ensuremath{\key_1}\xspace}
\newcommand{\prftwo}[1]{\ensuremath{\prf{#1}^2}\xspace}
\newcommand{\keytwo}{\ensuremath{\key_2}\xspace}
\newcommand{\answerkeytwo}{\ensuremath{\hat{\key}_2}\xspace}

\newcommand{\adversaries}{\ensuremath{A}\xspace}

\newcommand{\puzzles}{\ensuremath{P}\xspace}

\newcommand{\puzidx}{\ensuremath{p}\xspace}
\newcommand{\setsize}{\ensuremath{k}\xspace}

\newcommand{\totalsets}{\ensuremath{L}\xspace}

\newcommand{\indexset}{\ensuremath{I}\xspace}
\newcommand{\indexsetidx}{\ensuremath{\ell}\xspace}

\newcommand{\filesize}{\ensuremath{n}\xspace}

\newcommand{\FileOracle}{\ensuremath{\mathsf{content}}\xspace}
\newcommand{\HashOracle}{\ensuremath{\mathsf{hash}}\xspace}
\newcommand{\AnsOracle}{\ensuremath{\mathsf{ans}}\xspace}
\newcommand{\Queries}[1]{\ensuremath{q_{#1}}\xspace}

\newcommand{\Expt}{\ensuremath{\mathbf{Expt}}\xspace}
\newcommand{\BPAlgo}{\ensuremath{{\cal A}}\xspace}
\newcommand{\HashResult}{\ensuremath{h}\xspace}
\newcommand{\AnsResult}{\ensuremath{a}\xspace}
\newcommand{\secparam}{\ensuremath{\kappa}\xspace}
\newcommand{\String}{\ensuremath{\mathit{str}}\xspace}
\newcommand{\answerstring}{\ensuremath{\hat{\String}}\xspace}
\newcommand{\answerindexsetidx}{\ensuremath{\hat{\indexsetidx}}\xspace}
\newcommand{\challenge}{\ensuremath{\hat{\HashResult}}\xspace}
\newcommand{\answer}{\ensuremath{\hat{\AnsResult}}\xspace}
\newcommand{\confirm}{\ensuremath{\mathsf{confirm}}\xspace}

\newcommand{\E}[1]{\ensuremath{\mathsf{E}\left[{#1}\right]}}
\newcommand{\pr}[1]{\ensuremath{\mathsf{P}\left[{#1}\right]}}
\newcommand{\var}[1]{\ensuremath{\mathsf{Var}\left[{#1}\right]}}
\newcommand{\concat}{\ensuremath{||}}

\newcommand{\codertn}{\ensuremath{\mathbf{return}}\xspace}

\newcommand{\getsr}{{\:\stackrel{{\scriptscriptstyle R}}{\leftarrow}\:}}
\renewcommand{\gets}{{\:\leftarrow\:}}

\newtheorem{thm}{Theorem}[section]
\newtheorem{lem}[thm]{Lemma}

\newcounter{note}[section]

\newcommand{\totalHashNum}{\ensuremath{Q}}
\newcommand{\pCantSolv}{\ensuremath{p}}
\newcommand{\puzzlespercha}{\ensuremath{z}}

\author{Zhenghao Zhang \\
Computer Science Department \\ 
Florida State University, Tallahassee, FL, USA \\
zzhang@cs.fsu.edu}

\ignore{
\author{Zhenghao Zhang\thanks{Computer Science Department, Florida
State University, Tallahassee, FL, USA; {\tt zzhang@cs.fsu.edu}}
\and Vyas Sekar\thanks{Computer Science Department, Carnegie Mellon
University, Pittsburgh, PA, USA; {\tt vyass@cs.cmu.edu}}
\and  Michael K.\ Reiter\thanks{Department of Computer Science,
University of North Carolina, Chapel Hill, NC, USA;
{\tt reiter@cs.unc.edu}}
\and Hui Zhang\thanks{Computer Science Department, Carnegie Mellon
University, Pittsburgh, PA, USA; {\tt hzhang@cs.cmu.edu}}
}
}

\begin{document}

\title{A New Bound on the Performance of the Bandwidth Puzzle}

\pagestyle{plain}
\markboth{}{}

\maketitle

{\abstract 
A bandwidth puzzle was recently proposed to defend against colluding adversaries in peer-to-peer networks. The colluding adversaries do not do actual work but claim to have uploaded contents for each other to gain free credits from the system. The bandwidth puzzle guarantees that if the adversaries can solve the puzzle, they must have spent substantial bandwidth, the size of which is comparable to the size of the contents they claim to have uploaded for each other. Therefore, the puzzle discourages the collusion. In this paper, we study the performance of the bandwidth puzzle and give a lower bound on the {\em average} number of bits the adversaries must receive to be able to solve the puzzles with a certain probability. We show that our bound is tight in the sense that there exists a strategy to approach this lower bound asymptotically within a small factor. The new bound gives better security guarantees than the existing bound, and can be used to guide better choices of puzzle parameters to improve the system performance.}

\section{Introduction}

A key problem in peer-to-peer (p2p) based content sharing is the incentive for peers to contribute bandwidth to serve other peers \cite{peercontribution}.
Without a robust incentive mechanism, peers may choose not to upload contents for other peers, causing the entire system to fail. In many applications, a peer's contribution is measured by the number of bits it uploaded for other peers. It is difficult to measure the contribution because peers may collude with each other to get free credits. For example, if Alice and Bob are friends, Alice, without actually uploading, may claim that she has uploaded a certain amount of bits for Bob. Bob, when asked about this claim, will attest that it is true because he is Alice's friend. Therefore, Alice gets free credits. 

With the current Internet infrastructure, such collusions are difficult to detect, because the routers do not keep records of the traffic. Recently, a bandwidth puzzle scheme has been proposed solve this problem \cite{ourpuzzlepaper}. In the bandwidth puzzle scheme, a central credit manager, called the {\em verifier}, is assumed to exist in the network. The verifier issues puzzles to suspected nodes, called {\em provers}, to verify whether the claimed transactions are true. To be more specific, when the verifier suspects a set of provers for certain transactions, it issues puzzles {\em simultaneously} to all the involved provers, and asks them to send back answers within a time threshold. The puzzle's main features are (1) it takes time to solve a puzzle and (2) a puzzle can be solved only if the prover has access to the contents. To illustrate the basic idea of the puzzle, consider the previous simple example with Alice and Bob. The verifier issues two puzzles, one to Alice and one to Bob. As Alice did not upload the content for Bob, Alice has the content but Bob does not. When received the puzzles, Alice can solve hers and send the answer to the verifier before the threshold but not Bob. Bob also cannot ask help from Alice, because Alice cannot solve two puzzles within the threshold. Given this, Bob will fail to reply with the answer of the puzzle and the verifier will know that the transaction did not take place.

The bandwidth puzzle is most suited for live video broadcast applications, where fresh contents are generated constantly \cite{ourpuzzlepaper}. The verifier can naturally reside in the source node of the video, and the puzzle is based on the unique content currently being broadcast, such that there can be no existing contents downloaded earlier to solve the puzzles. The construction of bandwidth puzzle is simple and based only on hash functions and pseudorandom functions. In \cite{ourpuzzlepaper}, the puzzle scheme was implemented and incorporated into a p2p video distributing system, and was shown to be able to limit collusions significantly. An upper bound was also given for the expected number of puzzles that can be solved given the limit of the number of bits received among the adversaries. However, the bound  is ``loose in several respects,'' as stated by the authors, because its dominating term is quadratic to the number of adversaries such that it deteriorates quickly as the number of adversaries increases. In this paper, we give a much improved bound on the performance of the puzzle. The new bound gives the {\em average} number of bits the adversaries must have received if they can solve the puzzles with a certain probability. As we will prove, the average number of bits the adversaries receive is linear to the number of adversaries for all values of adversaries. It is also asymptotically tight, in the sense that there exists a strategy that achieves this bound asymptotically within a small factor. The improved bound leads to more relaxed constraints on the choice of puzzle parameters, which should in turn improve the system performance.

The rest of this paper is organized as follows. 
Section~\ref{sec:construction} describes the construction of the puzzle. 
Section~\ref{sec:proof} gives the proof of the new bound. 
Section~\ref{sec:implications} discusses the practical puzzle parameters and shows how a simple strategy approaches the bound.
Section~\ref{sec:relatedwork} discusses related works.
Section~\ref{sec:conclusions} concludes the paper.

\section{The Construction}
\label{sec:construction}

In this section, we describe the construction of the puzzle. The puzzle construction is largely the same as \cite{ourpuzzlepaper} except one difference: allowing repeated indices in one index set (the definition of index set will be given shortly), which simplifies the puzzle construction. We first give a high-level overview of the puzzle construction as well as introducing some notations. The main parameters of the puzzle are listed in Table \ref{table:listofnotations}.

\subsection{A High-level Description}

The content being challenged is referred to simply as {\em content}. There are $\filesize$ bits in the content, each given a unique index. An {\em index set} is defined as $\setsize$ ordered indices chosen from the $\filesize$ indices. Each index set defines a string denoted as $\String$, called the {\em true string} of this index set, which is obtained by reading the bits in the content according to the indices. $\String$ can be hashed using a hash function denoted as $\HashOracle$, and the output is referred to as the hash of the index set. To construct a puzzle, the verifier needs $\totalsets$ index sets denoted as $\indexset_1, \ldots, \indexset_{\totalsets}$, where an index set is obtained by randomly choosing the indices, allowing repeat. The verifier randomly chooses one index set among the $\totalsets$ index sets, denoted as $\indexset_{\answerindexsetidx}$, called the {\em answer index set}. It uses $\HashOracle$ to get the hash of $\indexset_{\answerindexsetidx}$, denoted as $\challenge$, which is called the {\em hint} of the puzzle. The puzzle is basically the $\totalsets$ index sets and $\challenge$. When challenged with a puzzle, the prover should prove that it knows which index set hashes into $\challenge$, by presenting another hash of $\indexset_{\answerindexsetidx}$ generated by hash function $\AnsOracle$. The purpose of using $\AnsOracle$ is to reduce the communication cost, as $\String_{\answerindexsetidx}$ may be long. The verifier may issue $\puzzlespercha$ puzzles to the prover and the prover has to solve the all puzzles before a time threshold $\threshold$. 

\begin{table}[t]
\begin{centering}
\begin{tabular}{|c|c|}
\hline 
$\filesize$ & The number of bits in the content \\
\hline 
$\setsize$ & The number of indices in an index set \\
\hline  
$\totalsets$  & The number of index sets in a puzzle \\
\hline 
$\puzzlespercha$  & The number of puzzles sent to a prover \\
\hline 
$\threshold$  & The time threshold to solve the puzzles \\
\hline 
\end{tabular}
\par\end{centering}
\caption{List of Puzzle Parameters}
\vspace{-0.4in}
\label{table:listofnotations} 
\end{table}

From a high level, the strengths of the puzzle are (1) a prover has to know the content, otherwise it cannot get the true strings of the index sets (2) even if the prover knows the content, it still needs to spend time to try different index sets until it finds an index set with the same hash as the hint, refereed to as a $\confirm$ event, because the hash function is one-way. In practice, the verifier need not generate all index sets; it need only generate and find the hash of the answer index set. The verifier should not send the $\totalsets$ index sets to the prover because this requires a large communication cost; instead, the verifier and the prover can agree on the same pseudorandom functions to generate the index sets and the verifier sends only a key for the pseudorandom functions. Therefore, this construction has low computation cost and low communication cost.

As a example, suppose $\filesize=8$ and the content is 00110101. Suppose $\setsize=4$, $\totalsets=3$, and the three index sets in the puzzle are  $\indexset_1 = \{5,3,7,0\}$, $\indexset_2 = \{1,2,6,3\}$, and $\indexset_3 = \{2,3,5,3\}$. Correspondingly, $\String_1=1110$, $\String_2=0101$ and $\String_3=1111$. Suppose the verifier chooses $\answerindexsetidx = 1$. Suppose $\HashOracle$ is the simply the the parity bit of the string, such that $\challenge = 1$. The prover receives the the hint and generates the three index sets, and finds that only $\indexset_1$ has parity bit 1. Suppose $\AnsOracle$ is simply the parity bit of every pair of adjacent bits. The prover presents `01' which proves that it knows  $\indexset_1$ is the answer index set.

\subsection{Detailed Puzzle Construction}

In the construction, it is assumed that the keys of the pseudorandom functions and the output of the hash functions are both $\secparam$ bits. In practice, $\secparam = 160$ suffices.

Pseudorandom functions are used to generate the index sets. A pseudorandom function family $\{\prf{\key}\}$ is a family of functions parameterized by a secret key. Roughly speaking, once initialized by a key, a pseudorandom function generates outputs that are indistinguishable from true random outputs. Two pseudorandom function families are used: $\{\prfone{\key}: \{1,\ldots,\totalsets\} \rightarrow \{0,1\}^\secparam\}$ and $\{\prftwo{\key}: \{1,\ldots,\setsize\} \rightarrow \{1, \ldots, \filesize\}\}$. 

Two hash functions are used in the construction, $\HashOracle$ and $\AnsOracle$.  $\HashOracle$ is used to get the hint. It actually hashes the concatenation of a $\secparam$-bit key, a number in the range of $[1,\totalsets]$, and a $\setsize$-bit string into $\secparam$-bits: $\{0,1\}^\secparam \times \{1, \ldots, \totalsets\} \times \{0,1\}^\setsize \rightarrow \{0,1\}^\secparam$. To prove the security of the puzzle,  $\HashOracle$ is modeled as a random oracle~\cite{bellare:RO}.  The other hash function is $\AnsOracle: \{0,1\}^\setsize \rightarrow \{0, 1\}^\secparam$. For $\AnsOracle$, only  collision-resistance is assumed.

As mentioned earlier, a puzzle consists of the hint $\challenge$ and $\totalsets$ index-sets. The verifier first randomly picks a $\secparam$-bit string as key $\keyone$. Then it randomly picks a number $\answerindexsetidx$ from $[1,\totalsets]$ as the index of the answer index set. With $\keyone$ and $\answerindexsetidx$, it generates $\keytwo^{\answerindexsetidx} \gets \prfone{\keyone}(\answerindexsetidx)$. $\keytwo^{\answerindexsetidx}$ is used as the key for $\prftwo{\keytwo}$ to generate the indices in the answer index set: $\indexset_{\answerindexsetidx} = \{\prftwo{\keytwo^{\answerindexsetidx}}(1)\ldots \prftwo{\keytwo^{\answerindexsetidx}}(\setsize)\}$. The verifier then finds $\String_{\answerindexsetidx}$. It then uses the concatenation of $\keyone$, $\answerindexsetidx$, and $\String_{\answerindexsetidx}$ as the input to $\HashOracle$ and uses the output as $\challenge$: $\challenge \gets \HashOracle(\keyone, \answerindexsetidx, \String_{\answerindexsetidx})$. Including $\keyone$ and $\answerindexsetidx$ ensures that the results of one puzzle-solving process cannot be used in the solving process of another puzzle, regardless of the content, $\setsize$, and $\totalsets$.  The prover can generate index sets in the same way as the verifier generates the answer index set, and can compare the hash of the index sets with the hint until a $\confirm$ is found. When the prover finds a $\confirm$ upon string $\String_{\indexsetidx}$, it returns $\AnsOracle(\String_{\indexsetidx})$.

\newcommand{\comment}[1]{}

\renewcommand{\thetable}{\arabic{table}}
\newtheorem{myth}{Theorem}
\def\comment#1{}
\newtheorem{defn}{Definition}
\newcommand{\B}{$\hfill$ \rule{2mm}{2mm}}
\setcounter{MaxMatrixCols}{15}
\newcommand{\DoubleSpace}{\edef\baselinestretch{1}\Large\normalsize}
\newcommand{\NewOracle}{\ensuremath{\Omega}}
\newcommand{\Hashrange}{\ensuremath{R}}
\newcommand{\Anshashrange}{\ensuremath{H}}
\newcommand{\SequenceIndex}{\ensuremath{\mathit{i}}}
\newcommand{\Sequence}{\ensuremath{\mathit{s}}}
\newcommand{\SequenceSet}{\ensuremath{\mathit{S}}}
\newcommand{\zAdvantage}{\ensuremath{\mathit{Adv}}}
\newcommand{\Hashoutput}{\ensuremath{\mathit{h}}}
\newcommand{\Hashoutputanswer}{\ensuremath{\mathit{h'}}}
\newcommand{\Bitsthreshold}{\ensuremath{\mathit{V}}}
\newcommand{\NewAlgo}{\ensuremath{{\cal B}}\xspace}
\newcommand{\Answerseq}{\ensuremath{\mathit{i^*}}}
\newcommand{\Answerhash}{\challenge}
\newcommand{\HashResultsecond}{\ensuremath{\mathit{h}}}
\newcommand{\Answerhashsecond}{\ensuremath{\mathit{h^*}}}
\newcommand{\Answerstring}{\ensuremath{\mathit{\IndSect^*}}}
\newcommand{\Result}{\ensuremath{\mathit{r}}}
\newcommand{\Prob}{\ensuremath{\mathbf{P}}}
\newcommand{\Filequeries}{\ensuremath{f}}
\newcommand{\Oraclequeries}{\ensuremath{\Queries{\HashOracle}}}
\newcommand{\AnsOraclequeries}{\ensuremath{y^*}}
\newcommand{\NewPuzzle}{\ensuremath{\mathit{newpuzz}}}
\newcommand{\Puzzle}{\ensuremath{\mathit{puzz}}}
\newcommand{\Sim}{\ensuremath{\mathit{Sim}}}
\newcommand{\pickedind}{\ensuremath{c}}
\newcommand{\constadv}{\ensuremath{\epsilon}}

\newcommand{\RecvBits}{\ensuremath{\mathit{\omega}}}

\newcommand{\SuccEvt}{\ensuremath{\mathit{C}}}
\newcommand{\randproc}{\ensuremath{\mathit{W}}}

\newcommand{\correct}{\ensuremath{\mathsf{correct}}}

\newcommand{\BitsRevd}{\ensuremath{\mathsf{RV}}}
\newcommand{\BitUnq}{\ensuremath{\mathsf{RP}}}
\newcommand{\newIndRV}{\ensuremath{Y}}
\newcommand{\newIndBRV}{\ensuremath{Z}}

\newcommand{\newindices}{\ensuremath{t}}

\newcommand{\NumPuzzles}{\ensuremath{\puzzles}}

\newcommand{\AdverAdv}{\ensuremath{\sigma }}
\newcommand{\NumRcvBits}{\ensuremath{w}}

\newcommand{\PuzzSet}{\ensuremath{Z}}
\newcommand{\IndexPuzz}{\ensuremath{j}}
\newcommand{\ResultSet}{\ensuremath{R}}
\newcommand{\IndexcAdvsr}{\ensuremath{v}}

\newcommand{\maxt}{\ensuremath{C}}
\newcommand{\ratiod}{\ensuremath{d}}

\newcommand{\IndSect}{\ensuremath{s}}
\newcommand{\cutoff}{\ensuremath{h}}
\newcommand{\inductind}{\ensuremath{j}}
\newcommand{\avgind}{\ensuremath{\mu}}
\newcommand{\pickseq}{\ensuremath{s}}
\newcommand{\adveraryidx}{\ensuremath{v}}
\newcommand{\numconsq}{\ensuremath{T}}
\newcommand{\numhashqmust}{\ensuremath{\beta}}
\newcommand{\nuqidcsavgq}{\ensuremath{U}}
\newcommand{\nuqidcsgivenq}{\ensuremath{u}}
\newcommand{\thmlpindidx}{\ensuremath{j}}
\newcommand{\rplcI}{\ensuremath{h}}
\newcommand{\cfgamma}{\ensuremath{a}}
\newcommand{\cfbeta}{\ensuremath{b}}
\newcommand{\optimal}{\ensuremath{*}}
\newcommand{\OPTobj}{\ensuremath{W}}
\newcommand{\shortfrac}{\ensuremath{\delta}}
\newcommand{\varspeed}{\ensuremath{x}}

\newcommand{\varNeedSam}{\ensuremath{Z}}
\newcommand{\varNeedSamVal}{\ensuremath{j}}
\newcommand{\varNeedSamInd}{\ensuremath{i}}
\newcommand{\sumNeedSam}{\ensuremath{S}}
\newcommand{\pNeedSam}{\ensuremath{p}}
\newcommand{\uniqx}{\ensuremath{\eta}}

\newcommand{\OBJ}{\ensuremath{\Psi}}

\newcommand{\numchosets}{\ensuremath{J}}

\section{The Security Bound} 
\label{sec:proof}

In this section, we derive the new bound for the bandwidth puzzle. Although the puzzle is designed to defend against colluding adversaries, we begin with the simple case when there is only one adversary given only one puzzle, because the proof for this simple case can be extended to the case when multiple adversaries are given multiple puzzles.

\begin{table}[t]
\begin{centering}
\begin{tabular}{|c|c|}
\hline 
$\Oraclequeries$ & The number of hash queries allowed, determined by $\threshold$\\
\hline 
$\NewOracle$  & A special oracle for hash and content queries\\
\hline 
$\Bitsthreshold$  & The maximum number of missed bits \\
\hline 
$\shortfrac$  & A positive number determined by puzzle parameters \\
\hline 
\end{tabular}
\par\end{centering}
\caption{List of Notations in the Proof}
\vspace{-0.4in}
\label{table:listofpara} 
\end{table}

\subsection{Single Adversary with a Single Puzzle} 

Consider a single adversary challenged with one puzzle. We begin with assumptions and definitions. Some key proof parameters and notations are listed in Table \ref{table:listofpara}.

\subsubsection{Assumptions and Definitions} In the proof, we model $\HashOracle$ and $\AnsOracle$ as random oracles and refer to them as the {\em hash oracle} and the {\em answer oracle}, respectively. Obtaining a bit in the content is also modeled as making a query to the {\em content oracle} denoted as $\FileOracle$. The adversary is given access to $\HashOracle$, $\AnsOracle$, and $\FileOracle$. To model the computational constraint of the prover in the limited time $\threshold$ allowed to solve the puzzle, we assume the number of queries to $\HashOracle$ is no more than $\Oraclequeries$. To ensure that honest provers can solve the puzzle, $\Oraclequeries \ge \totalsets$. However, we do not assume any limitations on the number of queries to $\FileOracle$ and $\AnsOracle$. We refer a query to $\FileOracle$ as a {\em content query} and a query to $\HashOracle$ a {\em hash query}. We use $\BPAlgo$ to denote the algorithm adopted by the adversary.

In our proof, we define a special oracle, $\NewOracle$, as an oracle that answers two kinds of queries, both the content query and the hash query. Let $\NewAlgo$ be an algorithm for solving the puzzle, when given access to the special oracle $\NewOracle$ and the answer oracle $\AnsOracle$. If $\NewAlgo$ makes a content query, $\NewOracle$ simply replies with the content bit. In addition, it keeps the history of the content queries made. We say a hash query to $\NewOracle$ is {\em informed} if there are no more than $\Bitsthreshold$ bits missing in the index set and {\em uninformed} otherwise, where $\Bitsthreshold$ is a proof parameter much smaller than $\setsize$. If $\NewAlgo$ makes an informed hash query for $\indexset_\indexsetidx$, $\NewOracle$ replies with the hash of $\indexset_\indexsetidx$; otherwise, it returns $\emptyset$. In addition, if $\NewAlgo$ makes more than $\totalsets$ hash queries for the puzzle, $\NewOracle$ will not answer further hash queries.

\subsubsection{Problem Formalization} The questions we seek to answer is: given $\Oraclequeries$, if the adversary has a certain advantage in solving the puzzle, how many content queries it must make to $\FileOracle$ {\em on average}? In the context of p2p content distribution, this is analogous to giving a lower bound on the average number of bits a peer must have downloaded if it can pass the puzzle challenge with a certain probability. Note that we emphasize on the average number of bits because a deterministic bound may be trivial: if the adversary happens to pick the answer index set in the first attempt of hash queries, only $\setsize$ content queries are needed. However, the adversary may be lucky once but unlikely to be always lucky. Therefore, if challenged with a large number of puzzles, the average number of queries it makes to $\FileOracle$ must be above a certain lower bound, which is the bound we seek to establish. 

In an earlier work \cite{ourpuzzlepaper}, an upper bound was given on the expected number puzzles that can be solved if the adversary is allowed $\Oraclequeries$ hash queries and a certain number of content queries. In this work, we remove assumption on the maximum number of content queries. With less assumptions, our proof is less restrictive and applies to more general cases. The new problem is different from the problem studied in \cite{ourpuzzlepaper}, and new techniques are needed to establish the bound. Note that although the adversaries is allowed to download as many bits as they wish, they prefer to employ an intelligent algorithm to minimize the number of downloaded bits because their intention is to use collusion to avoid spending bandwidth. The new bound guarantees that, if the adversaries wishes to have a certain advantage in solving the puzzles, there exists a lower bound on the average number of bits they have to download, regardless of the algorithm they adopt.

\subsubsection{Proof Sketch} A sketch of our proof is as follows. As it is difficult to derive the optimal algorithm the adversary may adopt, our proof is ``indirect.'' That is, by using $\NewOracle$, we introduce a simplified environment which is easier to reason about. We show that an algorithm can be found in the simplified environment with performance close to that of the best algorithm the adversary may adopt in the real environment. This provides a link between the simplified environment and the real environment: knowing the bound for the former, the bound for the latter is a constant away. We establish the performance bound of the optimal algorithm in the simplified environment, by showing that to solve the puzzle with certain probability, an algorithm must make a certain number of informed hash queries to $\NewOracle$ and the average number of unique indices in the informed queries, i.e., the number of content queries, is bounded.

\subsubsection{Proof Details}

\comment{
\begin{figure}
\begin{tabular}{|l|}
\hline
$\Expt(\NewAlgo)$: \\
~~~ $\FileOracle \getsr \funcs{\{1,\ldots,\filesize\}}{\{0,1\}}$ \\
~~~ $\HashOracle \getsr \funcs{\{0,1\}^\secparam \times \{1,\ldots,\totalsets\} \times \{0,1\}^\setsize}{\{0,1\}^\secparam}$\\
~~~ $\keyone \getsr \{0,1\}^\secparam$ \\
~~~ $\answerindexsetidx \getsr \{1, \ldots, \totalsets\}$ \\
~~~ $\answerkeytwo \gets \prfone{\keyone}(\answerindexsetidx)$ \\
~~~ $\answerstring \gets \FileOracle(\prftwo{\answerkeytwo}(1)) \concat \ldots \concat \FileOracle(\prftwo{\answerkeytwo}(\setsize))$ \\
~~~ $\challenge \gets \HashOracle(\keyone, \answerindexsetidx, \answerstring)$ \\
~~~ $\answer \gets \AnsOracle(\answerstring)$ \\
~~~ Clear the history in $\NewOracle$ \\
~~~ $\AnsResult \gets {\NewAlgo}^{\NewOracle,\AnsOracle}(\keyone,\challenge)$ \\
~~~ \If ($\AnsResult = \answer$) \\   
	~~~ ~~~\codertn 1 \\
~~~ \codertn 0 \\
\hline
\end{tabular}
\caption{Experiment for Theorem~\ref{thm:advantageGivenSpread}}
\label{fig:expt}
\end{figure}
}



Given any algorithm $\BPAlgo$ the adversaries may adopt, we construct an algorithm $\NewAlgo_{\BPAlgo}$ that employs $\BPAlgo$ and implements oracle queries for $\BPAlgo$. $\NewAlgo_{\BPAlgo}$ terminates when $\BPAlgo$ terminates, and returns what $\BPAlgo$ returns. When $\BPAlgo$ makes a query, $\NewAlgo_{\BPAlgo}$ replies as follows: 
\begin{algorithm}
  \caption{$\NewAlgo_{\BPAlgo}$ answers oracle queries for $\BPAlgo$}
  \label{algsingle}
  \begin{algorithmic}[1]
    \STATE When $\BPAlgo$ makes a query to $\FileOracle$, $\NewAlgo_{\BPAlgo}$ makes the same content query to $\NewOracle$ and {\bf returns} the result to $\BPAlgo$. 
	\STATE When $\BPAlgo$ makes a query to $\AnsOracle$, $\NewAlgo_{\BPAlgo}$ makes the same query to $\AnsOracle$ and {\bf returns} the result to $\BPAlgo$.  
	\STATE When $\BPAlgo$ makes a query to $\HashOracle$ for $\indexset_\indexsetidx$:

\begin{enumerate}
\item $\NewAlgo_{\BPAlgo}$ checks whether $\BPAlgo$ has made exactly the same query before. If yes, it {\bf returns} the same answer as the last time. 

\item $\NewAlgo_{\BPAlgo}$ checks whether there are no less than $\Bitsthreshold$ bits in $\indexset_\indexsetidx$ that have not been queried. If yes, it {\bf returns} a random string. 

\item $\NewAlgo_{\BPAlgo}$ checks whether it has made a hash query for $\indexset_\indexsetidx$ before. If no, $\NewAlgo_{\BPAlgo}$ makes a hash query to $\NewOracle$. If  $\confirm$ is obtained upon this query, $\NewAlgo_{\BPAlgo}$ knows that $\indexset_\indexsetidx$ is the answer index set, and sends content queries $\NewOracle$ to get the remaining bits in $\indexset_{\indexsetidx}$.

\item If $\indexset_\indexsetidx$ is not the answer index set, $\NewAlgo_{\BPAlgo}$ {\bf returns} a random string.  

\item If the string $\BPAlgo$ submitted is the true string of $\indexset_\indexsetidx$, $\NewAlgo_{\BPAlgo}$ {\bf returns} the hash of $\indexset_\indexsetidx$. 

\item $\NewAlgo_{\BPAlgo}$ {\bf returns} a random string.
\end{enumerate}

  \end{algorithmic}
  \label{greedy}
\end{algorithm}

Let $\RecvBits()$ denote the average number of bits received by an algorithm, where the average is taken over the random choices of the algorithm and the randomness of the puzzle. We have
\begin{thm}
Let $\SuccEvt_\BPAlgo$ be the event that $\BPAlgo$ returns the correct answer when $\BPAlgo$ is interacting directly with $\FileOracle$, $\HashOracle$ and $\AnsOracle$. Let $\SuccEvt_{\NewAlgo_{\BPAlgo}}$ be the event that $\NewAlgo_{\BPAlgo}$ returns the correct answer, when $\NewAlgo_{\BPAlgo}$ is interacting with $\NewOracle$ and $\AnsOracle$. Then, 
\[
\pr{\SuccEvt_{\NewAlgo_{\BPAlgo}}} \geq \pr{\SuccEvt_\BPAlgo} - \frac{\Oraclequeries}{2^{\Bitsthreshold}},
\] 
and 
\[
\RecvBits[\NewAlgo_{\BPAlgo}] \leq \RecvBits[\BPAlgo] + \frac{\totalsets \setsize \Oraclequeries}{2^{\Bitsthreshold}} + \Bitsthreshold.
\]
\label{thm_compadv}
\end{thm}

\begin{IEEEproof}
In our construction, $\NewAlgo_{\BPAlgo}$ employs $\BPAlgo$, and answers oracle queries for $\BPAlgo$. Denote the random process of $\BPAlgo$ when it is interacting directly with $\FileOracle$, $\HashOracle$ and $\AnsOracle$ as $\randproc$, and denote the random process of $\BPAlgo$ when it is interacting  with the oracles implemented by $\NewAlgo_{\BPAlgo}$ as $\randproc'$. We prove that $\randproc$ and $\randproc'$ will progress in the same way statistically with only one exception, while the probability of this exception is bounded.

First, we note that when $\BPAlgo$ makes a query to $\FileOracle$ or $\AnsOracle$, $\NewAlgo_{\BPAlgo}$ simply gives the query result, therefore the only case needs to be considered is when $\BPAlgo$ makes a query to $\HashOracle$. When $\BPAlgo$ makes a query for $\indexset_\indexsetidx$ to $\HashOracle$, 
\begin{itemize}
\item If there are still no less than $\Bitsthreshold$ unknown bits in this index set, $\NewAlgo_{\BPAlgo}$ will simply return a random string, which follows the same distribution as the output of the $\HashOracle$ modeled as a random oracle. If $\indexsetidx \neq \answerindexsetidx$, such a query will not result in a $\confirm$, and this will have same effect on the progress of the algorithm statistically as when $\BPAlgo$ is making a query to $\HashOracle$. However, if $\indexsetidx = \answerindexsetidx$, it could happen that $\BPAlgo$ is making a query with the true string. In this case, the exception occurs. That is, $\randproc'$ will not terminate, but $\randproc$ will terminate with the correct answer to the puzzle. However, the probability of this exception is bounded from the above by $\frac{\Oraclequeries}{2^{\Bitsthreshold}}$, because if no less than $\Bitsthreshold$ bits are unknown, the probability of making a hash query with the true string is no more than $\frac{\Oraclequeries}{2^{\Bitsthreshold}}$. 
\item If $\NewAlgo_{\BPAlgo}$ has made enough content queries for this index set, $\NewAlgo_{\BPAlgo}$ checks whether it has made hash query for this index set before. If no, $\NewAlgo_{\BPAlgo}$ makes the hash query, and if a $\confirm$ is obtained, $\NewAlgo_{\BPAlgo}$ knows that this is the answer index set and get the possible remaining bits in it; otherwise $\NewAlgo_{\BPAlgo}$ knows that it is not the answer index set. If $\indexset_\indexsetidx$ is not the answer index set, $\NewAlgo_{\BPAlgo}$ will simply return a random string, which will  have the same effect statistically on the progress of $\BPAlgo$ as when $\BPAlgo$ is interacting with $\HashOracle$. If $\indexset_\indexsetidx$ is the answer index set, $\NewAlgo_{\BPAlgo}$ checks whether $\BPAlgo$ is submitting the true string, and returns the true hash if yes and a random string otherwise. This, clearly, also has the same effect statistically of the progress of $\BPAlgo$ as when $\BPAlgo$ is interacting with $\HashOracle$.
\end{itemize} 

From the above discussion, we can see that $\pr{\SuccEvt_{\NewAlgo_{\BPAlgo}}}$ is no less than $\pr{\SuccEvt_\BPAlgo}$ minus the probability of the exception. Therefore, the first half of the theorem is proved. We can also see that if the exception occurs, $\NewAlgo_{\BPAlgo}$ makes at most $\totalsets \setsize$ more content queries than $\BPAlgo$. If the exception does not occur, $\NewAlgo_{\BPAlgo}$ receives at most $\Bitsthreshold$ bits than $\BPAlgo$ it encapsulates, and therefore at most $\Bitsthreshold$ bits more than $\BPAlgo$ on average when $\BPAlgo$ is interacting directly with $\FileOracle$, $\HashOracle$ and $\AnsOracle$. \end{IEEEproof}

Theorem \ref{thm_compadv} allows us to establish a connection between the ``real'' puzzle solver and the puzzle solver interacting with $\NewOracle$. The advantage of introducing $\NewOracle$ is that a good algorithm will not send any uninformed queries to $\NewOracle$, because it will get no information from such queries. If there is a bound on the number of hash queries, which are all informed, it is possible to establish a lower bound on the number of unique indices involved in such queries, with which the lower bound of the puzzle can be established. It is difficult to establish such bound based on $\HashOracle$ directly because $\HashOracle$ answers any queries. Although some queries are ``more informed'' than others, all queries have non-zero probabilities to get a $\confirm$. The next theorem establishes the lower bound on the expected number of informed hash queries to achieve a given advantage by an optimal algorithm interacting with $\NewOracle$. 
\begin{thm}
Suppose $\NewAlgo$ is an optimal algorithm for solving the puzzle when interacting with $\NewOracle$. If $\NewAlgo$ solves the puzzle with probability no less than $\constadv$, on average, the number of informed hash queries it makes is no less than 
$
\frac{(\constadv - \frac{1}{2^{\Bitsthreshold}}) (\totalsets + 1)}{2}. 
$
\label{thm_numinformedQ}
\end{thm}
\begin{IEEEproof}
Let $\correct$ denote the event that $\NewAlgo$ returns the correct answer. Note that
\begin{eqnarray}
\pr{\correct} &=& \pr{\correct \mid \confirm} \pr{\confirm} \nonumber \\
&& + \pr{\correct \mid \neg\confirm} \pr{\neg \confirm} \nonumber \\
&=& \pr{\confirm} \nonumber \\
&& + \pr{\correct \mid \neg\confirm} \pr{\neg \confirm} \nonumber \\
&\leq & \pr{\confirm} + \pr{\correct \mid \neg\confirm} \nonumber \\
&\leq & \pr{\confirm} + \frac{1}{2^{\Bitsthreshold}} \nonumber 
\end{eqnarray}
Note that $\pr{\correct \mid \neg\confirm} \le \frac{1}{2^{\Bitsthreshold}}$ because if the algorithm returns the correct answer, it must have the true string of the answer index set, since $\AnsOracle$ is collision-resistant. If a $\confirm$ was not obtained, the answer index set is missing no less than $\Bitsthreshold$ bits, since otherwise an optimal algorithm should make query which will result in a $\confirm$. Therefore, the probability that the algorithm can obtain the true string of the answer index set is no more than $\frac{1}{2^{\Bitsthreshold}}$. Note that hash queries to $\NewOracle$ will not help in the guessing of the true string, because  $\NewOracle$ is aware of the number of missing bits and will not reply with any information. Therefore, any algorithm that achieves advantage $\constadv$ in solving the puzzle must have an advantage of no less than $\constadv - \frac{1}{2^{\Bitsthreshold}}$ to get $\confirm$.

Let $P_1$ be the probability that $\NewAlgo$ makes no hash query and let $P_i$ be the probability that $\NewAlgo$ stops making hash queries after all previous queries (queries 1 to $i-1$) failed to generate a $\confirm$ for $2\le i \le \totalsets$. Consider the probability that a $\confirm$ is obtained upon the $i_{th}$ query. For a given set of $P_1, P_2, \ldots, P_{\totalsets}$, because $\answerindexsetidx$ is picked at random, the probability is 
\begin{eqnarray}
& & (1-P_1) \frac{\totalsets-1}{\totalsets} (1-P_2) \frac{\totalsets-2}{\totalsets-1} \ldots (1-P_{i}) \frac{1}{\totalsets-i}  \nonumber \\
&=& \frac{1}{\totalsets} \prod_{j=1}^{i} (1-P_j) \nonumber
\end{eqnarray}
Therefore, the probability that the algorithm can get a $\confirm$ is
\[\sum_{i=1}^{\totalsets}[ \frac{1}{\totalsets} \prod_{j=1}^{i} (1-P_j)]. \]

The event that exactly $i$ queries are made occurs when a $\confirm$ was obtained upon the $i_{th}$ query, or when all first $i$ queries failed to obtain the $\confirm$ and the algorithm decides to stop making queries. The probability is thus
\[[\prod_{j=1}^{i} (1-P_j)] [\frac{1}{\totalsets} + \frac{\totalsets - i}{\totalsets} P_{i+1}].\]
Note that $P_{L+1}$ is not previously defined. However, as $\frac{\totalsets - i}{\totalsets} = 0$ when $i = \totalsets$, for convenience, we can use the same expression for all  $1 \le i \le \totalsets$ for any arbitrary value of $P_{L+1}$. To derive the lower bound, we therefore need to solve the problem of minimizing 
\[\sum_{i=1}^{\totalsets} i [\prod_{j=1}^{i} (1-P_j)] [\frac{1}{\totalsets} + \frac{\totalsets - i}{\totalsets} P_{i+1}]\]
subject to constraint that 
\[ \sum_{i=1}^{\totalsets}[ \frac{1}{\totalsets} \prod_{j=1}^{i} (1-P_j)] = \constadv - \frac{1}{2^{\Bitsthreshold}}\]
and 
\[0 \leq P_i \leq 1.\]

To solve the problem, we let 
$
\eta_i = \prod_{j=1}^{i} (1-P_j)
$
and note that 
$
P_{i+1} = 1 - \frac{\eta_{i+1}}{\eta_{i}}.
$
Therefore, 
\begin{eqnarray}
& & \sum_{i=1}^{\totalsets} i [\prod_{j=1}^{i} (1-P_j)] [\frac{1}{\totalsets} + \frac{\totalsets - i}{\totalsets} P_{i+1}] \nonumber \\
&=& \sum_{i=1}^{\totalsets} i \eta_i [\frac{1}{\totalsets} + \frac{\totalsets - i}{\totalsets} (1 - \frac{\eta_{i+1}}{\eta_{i}})] \nonumber \\
&=& \frac{1}{\totalsets} [\sum_{i=1}^{\totalsets} (\totalsets - i + 1)\eta_i i - \sum_{i=1}^{\totalsets-1} (\totalsets - i)\eta_{i+1} i ] \nonumber \\
&=& \frac{1}{\totalsets} [\sum_{i=1}^{\totalsets} (\totalsets - i + 1) \eta_i ]. \nonumber
\end{eqnarray}

We therefore consider a new problem as minimizing 
$\frac{1}{\totalsets} [\sum_{i=1}^{\totalsets} (\totalsets - i + 1) \eta_i]$
subject to constraint that 
$\sum_{i=1}^{\totalsets} \eta_i = \totalsets (\constadv - \frac{1}{2^{\Bitsthreshold}})$, 
$0 \leq \eta_i \leq 1$,
$\eta_{i+1} \leq \eta_{i}$.  
The optimal value for the newly defined problem must be no more than that of the original problem, because any valid assignment of $\{P_{i}\}_{i}$ gives a valid assignment of $\{\eta_{i}\}_{i}$. To achieve the optimal value of the new problem, note that if $i<j$, the coefficient of $\eta_i$ is more than $\eta_j$ in the objective function, therefore, to minimize the objective function, we should reduce $\eta_i$ and increase $\eta_j$. 
Considering that $\{\eta_{i}\}_{i}$ is nondecreasing, the optimal is achieved when all $\eta_i$ are set to the same value $(\constadv - \frac{1}{2^{\Bitsthreshold}})$, and the optimal value is $\frac{(\constadv - \frac{1}{2^{\Bitsthreshold}}) (\totalsets + 1)}{2}$.
\end{IEEEproof}

Based on Theorem ~\ref{thm_numinformedQ}, any algorithm with an advantage of $\constadv$ must make no less than certain number of informed hash queries to $\NewOracle$ on average. We next derive the number of unique indices in a given number index sets. We first need the following lemma.
\begin{lem}
Suppose $\pickedind$ indices are randomly picked among $\filesize$ indices, with repeat. Let $\newIndRV$ be the random number denoting the number of unique indices among $\pickedind$ indices. Let $\avgind = \filesize (1-\shortfrac) [1-(1-\frac{1}{\filesize})^\pickedind]$ for a constant $0 < \shortfrac < 1$ and $\uniqx = \frac{\pickedind - \filesize \ln \frac{\filesize}{\filesize - \avgind}}{\filesize \sqrt{ (\frac{1}{\filesize - \avgind} - \frac{1}{\filesize})}}$. We have 
\[
\pr{\newIndRV \leq \avgind} \leq \frac{e^{-\uniqx^2/2}}{\sqrt{2\pi} \uniqx}\] 
\label{lem_couponcollector}
when $\filesize \rightarrow \infty$ and $\pickedind \rightarrow \infty$.
\end{lem}
\begin{IEEEproof}
Consider the process when indices are randomly taken from $\filesize$ indices. Let $\varNeedSam_\varNeedSamInd$ be random number denoting the number of samples needed to get the $\varNeedSamInd_{th}$ unique index. Clearly, $\pr{\varNeedSam_1 = 1} = 1$. In general, note that $\varNeedSam_\varNeedSamInd$ follows the geometric distribution, i.e., 
\[
\pr{\varNeedSam_\varNeedSamInd = \varNeedSamVal} = (\frac{\varNeedSamInd-1}{\filesize})^{\varNeedSamVal-1} \frac{\filesize - \varNeedSamInd+1}{\filesize}.
\]
Let $\pNeedSam_{\varNeedSamInd} = \frac{\filesize - \varNeedSamInd + 1}{\filesize}$, we have $\E{\varNeedSam_\varNeedSamInd} = \frac{1}{\pNeedSam_\varNeedSamInd}$, and $\var{\varNeedSam_\varNeedSamInd} = \frac{1-\pNeedSam_\varNeedSamInd}{\pNeedSam^2_\varNeedSamInd}$. Also, $\{\varNeedSam_\varNeedSamInd\}_{\varNeedSamInd}$ are independent of each other. Define $\sumNeedSam_\avgind = \sum_{\varNeedSamInd=1}^{\avgind} \varNeedSam_\varNeedSamInd$ and note that $\pr{\newIndRV \leq \avgind} = \pr{\sumNeedSam_\avgind \ge \pickedind}$. Therefore,  we will focus on finding $\pr{\sumNeedSam_\avgind \ge \pickedind}$.

Define $\varNeedSam'_\varNeedSamInd = \varNeedSam_\varNeedSamInd - \E{\varNeedSam_\varNeedSamInd}$. Let $\sumNeedSam'_\avgind = \sum_{\varNeedSamInd=1}^{\avgind} \varNeedSam'_\varNeedSamInd$ and note that $\pr{\sumNeedSam_\avgind \ge \pickedind} = \pr{\sumNeedSam'_\avgind \ge \pickedind - \sum_{\varNeedSamInd=1}^{\avgind} \frac{1}{\pNeedSam_\varNeedSamInd}}$. As $\{\varNeedSam'_\varNeedSamInd\}_{\varNeedSamInd}$ are independent random variables with zero mean, due to the Central Limit Theorem, $\sumNeedSam'_\avgind$ approximately follows the Gaussian distribution with zero mean and variance $\sum_{\varNeedSamInd=1}^{\avgind} \frac{1-\pNeedSam_\varNeedSamInd}{\pNeedSam^2_\varNeedSamInd}$. Note that as $\filesize \rightarrow \infty$ and $\pickedind \rightarrow \infty$, $\avgind \rightarrow \infty$, therefore, 
\[
\pr{\sumNeedSam'_\avgind \ge \pickedind - \sum_{\varNeedSamInd=1}^{\avgind} \frac{1}{\pNeedSam_\varNeedSamInd}} = Q(\frac{\pickedind - \sum_{\varNeedSamInd=1}^{\avgind} \frac{1}{\pNeedSam_\varNeedSamInd}}{\sqrt{\sum_{\varNeedSamInd=1}^{\avgind} \frac{1-\pNeedSam_\varNeedSamInd}{\pNeedSam^2_\varNeedSamInd}}}),
\]
where $Q()$ is the Gaussian Error Integral. To simplify the result, note that 
\[
\sum_{\varNeedSamInd=1}^{\avgind} \frac{1}{\pNeedSam_\varNeedSamInd} = \frac{\filesize}{\filesize} + \frac{\filesize}{\filesize - 1} + ... + \frac{\filesize}{\filesize -\avgind + 1}  \le \filesize \ln \frac{\filesize}{\filesize - \avgind}, 
\]
and 
\begin{eqnarray}
\sum_{\varNeedSamInd=1}^{\avgind} \frac{1-\pNeedSam_\varNeedSamInd}{\pNeedSam^2_\varNeedSamInd}  &=&   \sum_{\varNeedSamInd'=0}^{\avgind-1} \frac{1-\frac{\filesize - \varNeedSamInd'}{\filesize}}{(\frac{\filesize - \varNeedSamInd'}{\filesize})^2} 
= \sum_{\varNeedSamInd'=0}^{\avgind-1} \frac{\filesize \varNeedSamInd'}{(\filesize - \varNeedSamInd')^2} \nonumber \\
&=& \filesize \sum_{\varNeedSamInd'=0}^{\avgind-1} [\frac{\filesize}{(\filesize - \varNeedSamInd')^2} - \frac{\filesize - \varNeedSamInd'}{(\filesize - \varNeedSamInd')^2}] \nonumber \\
&<& \filesize^2 \sum_{\varNeedSamInd'=0}^{\avgind-1} \frac{1}{(\filesize - \varNeedSamInd')^2} \nonumber \\
&<& \filesize^2 (\frac{1}{\filesize - \avgind} - \frac{1}{\filesize}) \nonumber 
\end{eqnarray}
Applying these bounds, we have 
\[
\pr{\newIndRV \leq \avgind} \leq Q(\frac{\pickedind - \filesize \ln \frac{\filesize}{\filesize - \avgind}}{\filesize \sqrt{ (\frac{1}{\filesize - \avgind} - \frac{1}{\filesize})}})
\]
The proof completes since 
$
Q(x) \le \frac{e^{-x^2/2}}{\sqrt{2 \pi} x}
$ 
\cite{Qbound}.
\end{IEEEproof}

Note that according to the well-known coupon collector problem, $\filesize [1-(1-\frac{1}{\filesize})^\pickedind]$ is actually the average number of unique indices among $\pickedind$ indices, and $\shortfrac$ determines how far $\avgind$ deviates from this value. This lemma establishes the bound of the probability that the number of unique indices is less than $1-\shortfrac$ fraction of the average.

Let $\newIndRV^{\numchosets}_\pickseq$ denote the random variable of the minimum number of unique indices in $\pickseq$ index sets among all possible choices of $\pickseq$ index sets picked from $\numchosets$ index sets. Let $\avgind_\pickseq = \filesize (1-\shortfrac) [1-(1-\frac{1}{\filesize})^{\pickseq \setsize}]$ and $\uniqx_\pickseq = \frac{\pickseq \setsize - \filesize \ln \frac{\filesize}{\filesize - \avgind_\pickseq}}{\filesize \sqrt{ (\frac{1}{\filesize - \avgind_\pickseq} - \frac{1}{\filesize})}}$,
due to Lemma \ref{lem_couponcollector}, we have 
\[
\pr{\newIndRV^{\numchosets}_\pickseq \le \avgind_\pickseq} \leq \frac{e^{-\uniqx_\pickseq^2/2}}{\sqrt{2 \pi} \uniqx_\pickseq} \numchosets^\pickseq
\]
as $\filesize \rightarrow \infty$ and $\pickseq \setsize \rightarrow \infty$. This is because event $\newIndRV^{\numchosets}_\pickseq \le \avgind_\pickseq$ happens if one combination of $\pickseq$ index sets have no more than $\avgind_\pickseq$ unique indices, which happens with probability as given in Lemma \ref{lem_couponcollector}, and the total number of combinations to pick $\pickseq$ index sets is less than $\numchosets^\pickseq$. 

Considering practical puzzles, we note that $\filesize$ is very large, e.g., $10^7$, as well as $\setsize$, e.g., $10^4$. For any $\pickseq \ge 1$, as $\filesize \rightarrow \infty$ and $\pickseq \setsize \rightarrow \infty$, $(1-\frac{1}{\filesize})^{\pickseq \setsize}$ approaches $e^{- \pickseq \setsize / \filesize}$. We also pick puzzle parameters as well as $\shortfrac$, such that $\pr{\newIndRV^{\numchosets}_\pickseq \le \avgind_\pickseq}$ is negligibly small for conceivable values of $\numchosets$; see Section \ref{sec:implications} for details which has been confirmed by numerical analysis. Therefore, we guarantee that for the puzzles we use, {\em the number of unique indices in any $\pickseq$ index sets is no less than $(1-\shortfrac)\filesize (1-e^{- \pickseq \setsize / \filesize})$ with overwhelming probability.}

The next lemma is needed in determining the average number of unique indices in a certain average number of index sets.
\begin{lem}
Consider a linear programing problem of maximizing 
$
\sum_{i=0}^{\totalsets}  P_i e^{-i d}
$
subject to the constraint that 
$
\sum_{i=0}^{\totalsets}  P_i i = \numhashqmust
$,
$
\sum_{i=0}^{\totalsets}  P_i = \gamma
$, 
$
P_i \ge 0
$,
where $0 \le d$, $0 \le \gamma \le 1$ and $ 0 \le \numhashqmust \le \gamma \totalsets$. 
Denote the optimal value of the objective function as 
$F_{\totalsets}(\numhashqmust, \gamma)$, we have 
\[
F_{\totalsets}(\numhashqmust, \gamma) = \gamma-\frac{(1-e^{-\totalsets d})}{\totalsets} \numhashqmust 
\]
That is, to achieve the optimal value is to let $P_0 = \gamma-\frac{\numhashqmust}{\totalsets}$ and $P_\totalsets = \frac{\numhashqmust}{\totalsets}$, while let $P_i = 0$ for $0 < i < \totalsets$. 
\label{thm_linearprog}
\end{lem}
\begin{IEEEproof}
We will induction on $\totalsets$. To begin with, consider when $\totalsets = 1$. In this case, due to the constraints, $P_0$ and $P_1$ can be uniquely determined as $P_0 = \gamma - \numhashqmust$ and $P_1 = \numhashqmust$. Therefore, 
\[
F_1(\numhashqmust, \gamma)  =\gamma - \numhashqmust + \numhashqmust e^{-d},
\]
and the lemma is true. Suppose the lemma is true till $\thmlpindidx$. To get $F_{\thmlpindidx+1}(\numhashqmust, \gamma)$, suppose $P_{\thmlpindidx+1} = \gamma - \gamma'$, where $\gamma' \le \gamma$. Given this, $\numhashqmust' = \sum_{i=0}^{\thmlpindidx}  P_i i = \numhashqmust - (\gamma - \gamma') (\thmlpindidx+1)$, and therefore,
\begin{eqnarray}
F_{\thmlpindidx+1}(\numhashqmust, \gamma) &=& F_{\thmlpindidx}(\numhashqmust', \gamma') + (\gamma - \gamma')e^{-(\thmlpindidx+1)d}  \nonumber \\ 
	&=&  \gamma' - \frac{1-e^{-\thmlpindidx d}}{\thmlpindidx} \numhashqmust' + (\gamma - \gamma')e^{-(\thmlpindidx+1)d} \nonumber \\
	&=& \gamma' [1-\frac{(1-e^{-\thmlpindidx d}) (\thmlpindidx+1)}{\thmlpindidx} - e^{-(\thmlpindidx+1)d}] - \nonumber \\  
	&& \frac{1-e^{-\thmlpindidx d}}{\thmlpindidx} [\numhashqmust-\gamma(\thmlpindidx+1)] + \gamma e^{-(\thmlpindidx+1)d}\nonumber
\end{eqnarray}
Regarding $\gamma'$ as a variable, its coefficient is 
\[
1-\frac{(1-e^{-\thmlpindidx d}) (\thmlpindidx+1)}{\thmlpindidx} - e^{-(\thmlpindidx+1)d},
\] 
which is no more than 0. To see this, consider function 
\[
f(x) = 1-\frac{(1-e^{-\thmlpindidx x}) (\thmlpindidx+1)}{\thmlpindidx} - e^{-(\thmlpindidx+1)x}.
\] 
Note that $f(0) = 0$, and $f'(x) < 0$ when $x \ge 0$. Therefore, to maximize the objective function, $\gamma'$ should be as small as possible. Note that $\thmlpindidx P_{\thmlpindidx} = \numhashqmust - (\thmlpindidx+1) \gamma + (\thmlpindidx+1) \gamma'$ and $P_0 + P_{\thmlpindidx} = \gamma'$.
Therefore, 
\[
P_0=\frac{(\thmlpindidx+1)\gamma - \numhashqmust - \gamma'}{\thmlpindidx}.
\]
Since $P_0 \leq \gamma'$, we have $\gamma' \geq \gamma - \frac{\numhashqmust}{\thmlpindidx + 1}$. Therefore, $F_{\thmlpindidx+1}(\numhashqmust, \gamma) = \gamma - \frac{\numhashqmust}{\thmlpindidx+1} + \frac{\numhashqmust}{\thmlpindidx+1} e^{-(\thmlpindidx+1)d}$. 
\end{IEEEproof}

Suppose we randomly pick index sets from a total $\totalsets$ index sets when the average number of picked index sets is $\numhashqmust$. Let $P_i$ denote the probability that $i$ index sets are picked, where $0 \le i \le \totalsets$. We now give the lower bound of the average number of unique indices in the index sets picked, denoted as $\nuqidcsavgq(\numhashqmust)$. We have 
\[
\nuqidcsavgq(\numhashqmust) \ge \sum_{i=0}^{\totalsets} (1-\shortfrac)\filesize (1-e^{-i \setsize / \filesize}) P_i.
\]
Therefore, to derive the lower bound is to maximize
$\sum_{i=0}^{\totalsets}  P_i e^{-i \setsize / \filesize}$
subject to the constraints 
$\sum_{i=0}^{\totalsets}  P_i i = \numhashqmust$, 
$\sum_{i=0}^{\totalsets}  P_i = 1$,
$P_i \ge 0$. 

Based on Lemma \ref{thm_linearprog}, we immediately have 
\begin{lem}
If the average number of picked index sets is $\numhashqmust$, then
\[
\nuqidcsavgq(\numhashqmust) \ge  \frac{(1-\shortfrac)\filesize (1-e^{-\totalsets \setsize / \filesize}) \numhashqmust}{\totalsets}
\]
where $\shortfrac$ is a parameter determined by the puzzle parameters.
\label{thm_numuniqbits}
\end{lem}

\ignore{
We also have the following lemma, the proof of which is given in the accompanying technical report \cite{zztechrpt}.
\begin{lem}
Suppose an algorithm makes hash queries to $\numhashqmust$ index sets on average to solve the puzzle. Let $\nuqidcsavgq(\numhashqmust)$ be the average number of unique indices in the index sets selected by the algorithm. A constant $\shortfrac \in [0,1]$ exists and satisfies
\[
\nuqidcsavgq(\numhashqmust) \ge  \frac{(1-\shortfrac)\filesize (1-e^{-\totalsets \setsize / \filesize}) \numhashqmust}{\totalsets}.
\]
\label{thm_numuniqbits}
\end{lem}
Note that the lemma is trivially true if $\shortfrac = 1$; however, to make a useful bound, $\shortfrac$ should be as small as possible. We show in the accompanying technical report \cite{zztechrpt} that with practical puzzle parameters, $\shortfrac$ can be close to 0.
}

We may finally assemble the parts together. Suppose $\BPAlgo$ has an advantage of $\AdverAdv$ in solving the puzzle when receiving $\RecvBits(\BPAlgo)$ bits on average. Based on Theorem ~\ref{thm_compadv}, $\NewAlgo_{\BPAlgo}$ has an advantage of no less than $\AdverAdv-\frac{\Oraclequeries}{2^{\Bitsthreshold}}$ while receiving no more than $\RecvBits(\BPAlgo) + \frac{\totalsets \setsize \Oraclequeries}{2^{\Bitsthreshold}} + \Bitsthreshold$  bits on average. Based on Theorem ~\ref{thm_numinformedQ}, to achieve an advantage of at least $\AdverAdv-\frac{\Oraclequeries}{2^{\Bitsthreshold}}$, an algorithm must make at least $\frac{(\AdverAdv - \frac{\Oraclequeries+1}{2^{\Bitsthreshold}}) (\totalsets + 1)}{2}$ hash queries. Based on Lemma \ref{thm_numuniqbits}, also considering that $\NewAlgo$ needs to receive only $\setsize - \Bitsthreshold + 1$ bits per index set, $\NewAlgo$ receives at least $\nuqidcsavgq(\frac{(\AdverAdv - \frac{\Oraclequeries+1}{2^{\Bitsthreshold}}) (\totalsets + 1)}{2}) - \totalsets (\Bitsthreshold-1)$ bits on average. Therefore,  
\begin{thm}
Suppose $\BPAlgo$ solves the puzzle with probability no less than $\AdverAdv$. Let $\RecvBits(\BPAlgo)$ denote the average number of received bits. We have 
\begin{eqnarray}
\RecvBits(\BPAlgo) 
&\ge& \frac{(1-\shortfrac)\filesize (1-e^{-\totalsets \setsize / \filesize}) (\AdverAdv - \frac{\Oraclequeries+1}{2^{\Bitsthreshold}}) (\totalsets + 1)}{2 \totalsets} \nonumber \\
&& - \totalsets (\Bitsthreshold - 1)  - \frac{\totalsets \setsize \Oraclequeries}{2^{\Bitsthreshold}} - \Bitsthreshold \nonumber 
\end{eqnarray}
where $\Oraclequeries$, $\Bitsthreshold$, and $\shortfrac$ are constants determined by the puzzle parameters.
\end{thm}

%
%
%
%
%
%
%
%
%
%
%
%
%
%
%
%
%
%
%
%
%
%
%
%
%
%
%
%
%
%
%
%
%
%
%
%
%
%
%
%
%
%
%
%
%

\subsection{Multiple Adversaries with Multiple Puzzles}

We next consider the more complicated case when multiple adversaries are required to solve multiple puzzles. Suppose there are $\adversaries$ adversaries, and the number of puzzles they attempt to solve is $\NumPuzzles$. Note that $\NumPuzzles$ is greater than $\adversaries$ when $\puzzlespercha > 1$.

\subsubsection{Proof Sketch}
The proof uses the same idea as the single adversary case. Basically, we extend $\NewOracle$ to handle multiple adversaries, where $\NewOracle$ gives correct answer to a hash query from an adversary only if the number of bits the adversary received for the index set is greater than $\setsize - \Bitsthreshold$, {\em regardless of the number of bits other adversaries received}. With similar arguments as the single adversary case, we can establish the relationship between the algorithm performance when interacting with $\NewOracle$ and with the real oracles. We also obtain the average number of informed queries the adversaries must make to achieve certain advantages when interacting with $\NewOracle$. The bound is established after solving several optimization problems.

\subsubsection{Proof Details}
Suppose the adversaries run an algorithm $\BPAlgo$ that solves the $\NumPuzzles$ puzzles with probability $\AdverAdv$ while receiving $\RecvBits(\BPAlgo)$ bits on average. We wish to bound from below $\RecvBits(\BPAlgo)$ for a given $\AdverAdv$. We extend the definition of $\NewOracle$ and let it remember the content queries from each adversary. We use $\indexset^\puzidx_\indexsetidx$ to denote index set $\indexsetidx$ in puzzle $\puzidx$ where $1 \le \puzidx \le \NumPuzzles$. If an adversary makes a hash query for $\indexset^\puzidx_\indexsetidx$ while this adversary has made content query for more than $\setsize - \Bitsthreshold$ bits in $\indexset^\puzidx_\indexsetidx$,  $\NewOracle$ replies with the hash of $\indexset^\puzidx_\indexsetidx$, otherwise, it returns $\emptyset$. In addition, if $\NewAlgo$ makes more than $\totalsets$ hash queries for a particular puzzle, $\NewOracle$ will not answer further hash queries for this puzzle.

Similar to the single puzzle case, given an algorithm $\BPAlgo$ for solving the puzzles, we construct an algorithm $\NewAlgo_{\BPAlgo}$ employing $\BPAlgo$ denoted as $\NewAlgo_{\BPAlgo}$. $\NewAlgo_{\BPAlgo}$ terminates when $\BPAlgo$ terminates, and returns what $\BPAlgo$ returns. Algorithm \ref{alg:multiple} describes how $\NewAlgo_{\BPAlgo}$ implements oracle queries for $\BPAlgo$ which is very similar to the single adversary case.
\begin{algorithm} 
  \caption{$\NewAlgo_{\BPAlgo}$ answers oracle queries for $\BPAlgo$}
  \label{algsingle}
  \begin{algorithmic}[1]
    \STATE When $\BPAlgo$ makes a query to $\FileOracle$, $\NewAlgo_{\BPAlgo}$ makes the same content query to $\NewOracle$ and gives the result to $\BPAlgo$. 
	\STATE When $\BPAlgo$ makes a query to $\AnsOracle$, $\NewAlgo_{\BPAlgo}$ makes the same query to $\AnsOracle$ and gives the result to $\BPAlgo$.  
	\STATE When $\BPAlgo$ makes a query for $\indexset^\puzidx_\indexsetidx$ to $\HashOracle$ at adversary $\adveraryidx$:

\begin{enumerate}
\item $\NewAlgo_{\BPAlgo}$ checks whether adversary $\adveraryidx$ has made exactly the same query before. If yes, it {\bf returns} the same answer.

\item $\NewAlgo_{\BPAlgo}$ checks whether there are no less than $\Bitsthreshold$ bits in $\indexset^\puzidx_\indexsetidx$ that have not been queried for at adversary $\adveraryidx$, and if this is true, it {\bf returns} a random string. 

\item $\NewAlgo_{\BPAlgo}$ checks if it has made a hash query for $\indexset^\puzidx_\indexsetidx$, if no, it makes a hash query to $\NewOracle$. If  
$\confirm$ 
is obtained upon this query, $\NewAlgo_{\BPAlgo}$ knows $\indexset^\puzidx_\indexsetidx$ is the answer index set of puzzle $\puzidx$. $\NewAlgo_{\BPAlgo}$ sends content queries $\NewOracle$ to get the remaining bits in $\indexset^\puzidx_{\indexsetidx}$.

\item If $\indexset^\puzidx_\indexsetidx$ is not the answer index set of puzzle $\puzidx$, $\NewAlgo_{\BPAlgo}$ {\bf returns} a random string. 

\item If the string $\BPAlgo$ submitted is the true string of $\indexset^\puzidx_\indexsetidx$, $\NewAlgo_{\BPAlgo}$ {\bf returns} the hash of $\indexset^\puzidx_\indexsetidx$. 

\item  $\NewAlgo_{\BPAlgo}$ {\bf returns} a random string. 
\end{enumerate}

  \end{algorithmic}
  \label{alg:multiple}
\end{algorithm}

With very similar arguments as in Theorem \ref{thm_compadv}, we can have
\begin{thm}
Let $\SuccEvt_\BPAlgo$ be the event that $\BPAlgo$ returns the correct answers when it is interacting directly with $\FileOracle$, $\HashOracle$ and $\AnsOracle$. Let $\SuccEvt_{\NewAlgo_{\BPAlgo}}$ be the event that $\NewAlgo_{\BPAlgo}$ returns the correct answers, when it is interacting with $\NewOracle$ and $\AnsOracle$. Then, 
\[
\pr{\SuccEvt_{\NewAlgo_{\BPAlgo}}} \geq \pr{\SuccEvt_\BPAlgo} - \frac{\adversaries \Oraclequeries}{2^{\Bitsthreshold}},
\] 
and 
\[
\RecvBits[\NewAlgo_{\BPAlgo}] \leq \RecvBits[\BPAlgo] + \frac{\NumPuzzles \totalsets \setsize \adversaries \Oraclequeries}{2^{\Bitsthreshold}} + \Bitsthreshold \NumPuzzles.
\]
\label{thm_compadvmulti}
\end{thm}

Let $\NewAlgo$ denote the optimal algorithm for solving the puzzles when the algorithm is interacting with $\NewOracle$. Note that if the probability that $\NewAlgo$ solves all puzzles is no less than $\constadv$, the probability that an individual puzzle is solved is no less than $\constadv$. Based on Theorem \ref{thm_numinformedQ}, if a puzzle is solved with probability no less than $\constadv$, the average number of hash queries made for this puzzle is no less than $\frac{(\constadv - \frac{1}{2^{\Bitsthreshold}})(\totalsets+1)}{2}$. There are $\NumPuzzles$ puzzles, and we obtain the following theorem due to the linearity of expectation. 
\begin{thm}
If the probability that $\NewAlgo$ solves all puzzles is no less than $\constadv$, on average, the number of informed hash queries is no less than 
\[
\frac{\NumPuzzles (\constadv - \frac{1}{2^{\Bitsthreshold}})(\totalsets+1)}{2}
\]
\label{thm_numinformedQmulti}
\end{thm}

Next, we wish to bound from below the number of unique indices if the adversaries collectively have to make $\numconsq$ informed hash queries. Here we define the unique indices at adversary $\IndexcAdvsr$ as the total number of unique indices in the index sets that it made hash queries for, and denote it as $\nuqidcsgivenq_\IndexcAdvsr$. The total number of unique indices is defined as $\sum_{\IndexcAdvsr=1}^{\adversaries} \nuqidcsgivenq_\IndexcAdvsr$. Recall that  $\NewOracle$ will not answer a hash query from adversary $\IndexcAdvsr$ if adversary $\IndexcAdvsr$ has not received enough number bits for this index set. Note that $\NewOracle$ will not answer the hash query even if there exists another adversary knowing  enough bits for this index set. In other words, content queries made at one adversary do not count as content queries at other adversaries, which is the one of the key differences between the single adversary case and the multiple adversary case. $\NewAlgo$ may be able to assign hash queries to the adversaries intelligently, such that $\sum_{\IndexcAdvsr=1}^{\adversaries} \nuqidcsgivenq_\IndexcAdvsr$ is minimized. For instance, if two index sets share a large number of indices, they should be assigned to the same adversary. Nevertheless, we have
\begin{lem}
If the number of informed queries made by $\NewAlgo$ is $\numconsq$, 
\[
\sum_{\IndexcAdvsr=1}^{\adversaries} \nuqidcsgivenq_\IndexcAdvsr \ge (1-\shortfrac)\filesize [t(1-e^{-\Oraclequeries \setsize / \filesize}) + (1-e^{-(\numconsq - \Oraclequeries t) \setsize / \filesize})],
\] 
where 
$t = \mid \numconsq / \Oraclequeries \mid^-$ and $\mid x \mid^-$ denotes the largest integer no more than $x$.
\label{lem_multiadvindex}
\end{lem}
\begin{IEEEproof}
An adversary may make no more than $\Oraclequeries$ queries. Suppose the number of hash queries made by adversary $\IndexcAdvsr$ is $\pickseq_{\IndexcAdvsr}$. We have
\[
\sum_{\IndexcAdvsr=1}^{\adversaries} \nuqidcsgivenq_\IndexcAdvsr \ge \sum_{\IndexcAdvsr=1}^{\adversaries}(1-\shortfrac)\filesize (1-e^{- \pickseq_{\IndexcAdvsr}\setsize / \filesize})
\]
Therefore, to minimize $\sum_{\IndexcAdvsr=1}^{\adversaries} \nuqidcsgivenq_\IndexcAdvsr$ is to maximize 
$\sum_{\IndexcAdvsr=1}^{\adversaries} e^{-\pickseq_{\IndexcAdvsr}\setsize / \filesize}$
subject to the constraints that 
$\sum_{\IndexcAdvsr=1}^{\adversaries} \pickseq_{\IndexcAdvsr} = \numconsq$
and 
$0 \le \pickseq_{\IndexcAdvsr} \leq \Oraclequeries$.

We claim that the optimal is achieved when $\pickseq_i$ is set to be $\Queries{\HashOracle}$ for $1 \le i \le \mid \numconsq / \Oraclequeries \mid^-$, which we show by induction on the number of adversaries. First consider when $\adversaries = 2$. If $\numconsq \leq \Queries{\HashOracle}$, we claim that 
$\sum_{i=1}^{2} e^{-\pickseq_i \setsize / \filesize}$ is maximized when $\pickseq_1 = \numconsq$ and $\pickseq_2 = 0$, which is because for any valid $\pickseq_1$ and $\pickseq_2$,
\begin{eqnarray}
&& (1+e^{-\numconsq \setsize / \filesize}) - (e^{-\pickseq_1 \setsize / \filesize} + e^{-\pickseq_2 \setsize / \filesize}) \nonumber \\
&=& (1 - e^{-\pickseq_1 \setsize / \filesize}) (1 - e^{-\pickseq_2 \setsize / \filesize}) \nonumber \\
&\ge & 0. \nonumber 
\end{eqnarray}
Similarly, if $\Queries{\HashOracle} \leq  \numconsq \leq 2 \Queries{\HashOracle}$, $\sum_{i=1}^{\adversaries} e^{-\pickseq_i \setsize / \filesize}$ is maximized when $\pickseq_1 = \Queries{\HashOracle}$ and $\pickseq_2 = \numconsq - \Queries{\HashOracle}$, which is because for any valid $\pickseq_1$ and $\pickseq_2$,
\begin{eqnarray}
&&(e^{- \Queries{\HashOracle} \setsize / \filesize}+e^{-(\numconsq - \Queries{\HashOracle})\setsize / \filesize}) - (e^{-\pickseq_1 \setsize / \filesize} + e^{-\pickseq_2 \setsize / \filesize}) \nonumber \\
&=& (1 - e^{(-\pickseq_1 + \numconsq - \Queries{\HashOracle}) \setsize / \filesize}) (e^{(- \numconsq +\Queries{\HashOracle}) \setsize / \filesize} - e^{- \pickseq_2 \setsize / \filesize}) \nonumber \\
&\ge & 0. \nonumber 
\end{eqnarray}
Therefore our claim is true for $\adversaries = 2$. Suppose our claim is true for $\adversaries = \inductind$. For $\adversaries = \inductind + 1$, suppose in the optimal assignment, $\pickseq_{\inductind + 1} = 0$. Then, our claim is true based on the induction hypothesis. If in the optimal assignment, $\pickseq_{\inductind + 1} > 0$, we prove that in the optimal assignment $\pickseq_{i} = \Queries{\HashOracle}$ for all $1 \leq i \leq \inductind$, therefore our claim is still true. This because if $\pickseq_{i} < \Queries{\HashOracle}$ for some $i$, we can increase $\pickseq_{i}$ while decreasing $\pickseq_{\inductind + 1}$. Using similar arguments as for case when $\adversaries = 2$, this will increase the objective function, thus violating the fact that the assignment is optimal. 
\end{IEEEproof}

Similar to the single adversary case, if the average number of informed hash queries $\NewAlgo$ makes is $\numhashqmust$, we need to bound from below the average number of unique indices in the involved index sets, denoted as $\nuqidcsavgq(\numhashqmust)$. 
\begin{lem}
Consider there are $\adversaries$ adversaries given $\NumPuzzles$ puzzles, if the average number of informed hash queries is $\numhashqmust$, 
\[
\nuqidcsavgq(\numhashqmust) \ge  \frac{(1-\shortfrac)\filesize \numhashqmust (1-e^{-\Oraclequeries \setsize / \filesize})}{\Oraclequeries}
\]
\label{thm_multiadvindexavg}
\end{lem}
\begin{IEEEproof}
Denote the probability that there are $i$ queries as $P_i$, where $0 \le i \le \NumPuzzles \totalsets$. For notational simplicity, in this proof, we let $\ratiod = \setsize / \filesize$. Based on Lemma \ref{lem_multiadvindex}, we want to bound from below 
\[
\sum_{i=0}^{\NumPuzzles \totalsets} P_i[(t_i-1)(1-e^{-\Oraclequeries \ratiod}) + (1-e^{-[i - \Oraclequeries (t_i-1)] \ratiod})]
\]
under the constraints that 
\begin{eqnarray}
\sum_{i=0}^{\NumPuzzles \totalsets} P_i i = \numhashqmust, \nonumber \\
\sum_{i=0}^{\NumPuzzles \totalsets} P_i = 1, \nonumber \\
P_i \ge 0, \nonumber
\end{eqnarray}
where $t_i$ is an integer such that $(t_i - 1) \Oraclequeries \leq i < t_i \Oraclequeries$. To solve this problem, suppose $\maxt$ is the minimum integer satisfying $\NumPuzzles \totalsets \leq \maxt \Oraclequeries-1$, we relax the problem to minimizing
\[
\OBJ = \sum_{i=0}^{\maxt \Oraclequeries-1} P_i[(t_i-1)(1-e^{-\Oraclequeries \ratiod}) + (1-e^{-[i - \Oraclequeries (t_i-1)] \ratiod})]
\]
under the same constraints that 
\begin{eqnarray}
\sum_{i=0}^{\maxt \Oraclequeries-1} P_i i = \numhashqmust, \nonumber \\
\sum_{i=0}^{\maxt \Oraclequeries-1} P_i = 1, \nonumber \\
P_i \ge 0. \nonumber 
\end{eqnarray}

The optimal of relaxed problem will be no more than the optimal of the original problem. To solve the relaxed problem, let 
\[
\OBJ_{\IndSect} = \sum_{i=(\IndSect-1)\Oraclequeries }^{\IndSect \Oraclequeries - 1} P_i[(\IndSect-1)(1-e^{-\Oraclequeries \ratiod}) + (1-e^{-[i - \Oraclequeries (\IndSect-1)] \ratiod})]
\]
for $1 \leq  \IndSect \leq \maxt$. Clearly, $\OBJ = \sum_{\IndSect=1}^{\maxt} \OBJ_{\IndSect}$. 

Suppose a set of $\{\gamma_{\IndSect}\}_{\IndSect}$ and $\{\numhashqmust_{\IndSect}\}_{\IndSect}$ are given where $\sum_{\IndSect=1}^{\maxt} \gamma_{\IndSect} = 1$ and $\sum_{\IndSect=1}^{\maxt} \numhashqmust_{\IndSect} = \numhashqmust$. The set of $\{\gamma_{\IndSect}\}_{\IndSect}$ and $\{\numhashqmust_{\IndSect}\}_{\IndSect}$ are called {\em feasible} if it is possible to find $\{P_i\}_i$ such that 
\[
\sum_{i=(\IndSect-1)\Oraclequeries}^{\IndSect \Oraclequeries-1} P_i = \gamma_{\IndSect},
\]
and 
\[
\sum_{i=(\IndSect-1)\Oraclequeries}^{\IndSect \Oraclequeries-1} P_i i= \numhashqmust_{\IndSect}
\]
for all $1 \le  \IndSect \le \maxt$. Note that $\{\gamma_{\IndSect}\}_{\IndSect}$ and $\{\numhashqmust_{\IndSect}\}_{\IndSect}$ are feasible if and only if 
\[
(\IndSect-1) \Oraclequeries \gamma_{\IndSect} \le \numhashqmust_{\IndSect} \le (\IndSect \Oraclequeries-1)\gamma_{\IndSect}.
\]

When $\{\gamma_{\IndSect}\}_{\IndSect}$ and $\{\numhashqmust_{\IndSect}\}_{\IndSect}$ are given and are feasible, to minimize $\OBJ$ is to minimize each individual $\OBJ_{\IndSect}$. Note that 
\begin{eqnarray}
\OBJ_\IndSect &=& \gamma_{\IndSect} [(\IndSect-1)(1-e^{-\Oraclequeries \ratiod}) + 1] \nonumber \\
&& - \sum_{i=(\IndSect-1)\Oraclequeries}^{\IndSect \Oraclequeries-1} P_i e^{-[i - \Oraclequeries (\IndSect-1)] \ratiod} \nonumber \\
&=& \gamma_{\IndSect} [(\IndSect-1)(1-e^{-\Oraclequeries \ratiod}) + 1] \nonumber \\
&&- \sum_{\rplcI=0}^{\Oraclequeries-1} P_{\Oraclequeries (\IndSect-1) + \rplcI} e^{-\rplcI \ratiod}, \nonumber
\end{eqnarray}
where $\rplcI = i - \Oraclequeries (\IndSect-1)$. Note that if 
\[
\sum_{i=(\IndSect-1)\Oraclequeries}^{\IndSect \Oraclequeries-1} P_i i= \numhashqmust_{\IndSect},
\]
then 
\[
\sum_{\rplcI=0}^{\Oraclequeries-1} P_{\Oraclequeries (\IndSect-1) + \rplcI} \rplcI= \numhashqmust_{\IndSect} - (\IndSect - 1) \Oraclequeries \gamma_\IndSect.
\] 
Denote the minimum value of $\OBJ_{\IndSect}$ for given $\gamma_{\IndSect}$ and $\numhashqmust_{\IndSect}$ as $\OBJ^{\gamma_{\IndSect},\numhashqmust_{\IndSect}}_\IndSect$. Applying Lemma \ref{thm_linearprog},
\begin{eqnarray}
\OBJ^{\gamma_{\IndSect},\numhashqmust_{\IndSect}}_\IndSect &=& \gamma_{\IndSect} [(\IndSect-1)(1-e^{-\Oraclequeries \ratiod}) + 1] - \gamma_\IndSect \nonumber \\
&& + [\frac{1-e^{-(\Oraclequeries - 1) \ratiod}}{\Oraclequeries - 1}] [\numhashqmust_{\IndSect} - (\IndSect - 1)\Oraclequeries \gamma_\IndSect]\nonumber \\
&=& \gamma_{\IndSect} (\IndSect-1)[ \frac{\Oraclequeries e^{-(\Oraclequeries-1 ) \ratiod}}{\Oraclequeries - 1} - e^{-\Oraclequeries \ratiod}  \nonumber  \\
&& - \frac{1}{\Oraclequeries-1}]  + [\frac{1-e^{-(\Oraclequeries - 1) \ratiod}}{\Oraclequeries - 1}] \numhashqmust_{\IndSect} \nonumber
\end{eqnarray}
Let 
\[
\cfgamma = \frac{\Oraclequeries e^{-(\Oraclequeries-1 ) \ratiod}}{\Oraclequeries - 1} - e^{-\Oraclequeries \ratiod} - \frac{1}{\Oraclequeries-1}
\] 
and 
\[
\cfbeta = \frac{1-e^{-(\Oraclequeries - 1) \ratiod}}{\Oraclequeries - 1},
\]
we have 
\[
\OBJ \ge \sum_{\IndSect=1}^{\maxt} \OBJ^{\gamma_{\IndSect},\numhashqmust_{\IndSect}}_\IndSect = \cfbeta \numhashqmust + \cfgamma \sum_{\IndSect=1}^{\maxt} (\IndSect-1) \gamma_\IndSect.
\]
We also note that $\cfgamma<0$, which is because function 
\[
f(x)=\frac{\Oraclequeries e^{-(\Oraclequeries-1 ) x}}{\Oraclequeries - 1} - e^{-\Oraclequeries x} - \frac{1}{\Oraclequeries-1}
\] 
is 0 when $x=0$, while $f'(x) < 0$ for $x > 0$. Therefore, finding the minimum value of $\OBJ$ is equivalent to finding a set of feasible $\{\gamma_{\IndSect}\}_{\IndSect}$ and $\{\numhashqmust_{\IndSect}\}_{\IndSect}$ such that $\sum_{\IndSect=1}^{\maxt} (\IndSect-1) \gamma_\IndSect$ is maximized.

We consider the problem of maximizing 
\[
\OPTobj = \sum_{\IndSect=1}^{\maxt} (\IndSect-1) \gamma_\IndSect
\] 
subject to the constraints that 
\[
(\IndSect-1) \Oraclequeries \gamma_{\IndSect} \le \numhashqmust_{\IndSect} \le (\IndSect \Oraclequeries-1)\gamma_{\IndSect},
\]
\[
\sum_{\IndSect=1}^{\maxt} \gamma_{\IndSect} = \gamma,
\] 
\[
\sum_{\IndSect=1}^{\maxt} \numhashqmust_{\IndSect} = \numhashqmust,
\]
where $0 \le \gamma \le 1$ and $0 \le  \numhashqmust \le \gamma (\maxt \Oraclequeries - 1)$. Denote the maximum value of $\OPTobj$ as $\OPTobj^\optimal$. We claim that 
\begin{itemize}
\item If $(\maxt-1)\Oraclequeries \gamma <  \numhashqmust $, $\OPTobj^\optimal = \gamma (\maxt-1)$ and the optimal is achieved when $\gamma_\maxt = \gamma$, $\numhashqmust_\maxt = \numhashqmust$, while $\gamma_\IndSect = 0$ and $\numhashqmust_\IndSect = 0$ for $1 \le \IndSect < \maxt$; 
\item If $(\maxt-1)\Oraclequeries \gamma \ge  \numhashqmust $, $\OPTobj^\optimal = \frac{\numhashqmust}{\Oraclequeries}$, and the optimal value is achieved when $\gamma_1 = \gamma - \frac{\numhashqmust}{(\maxt-1)\Oraclequeries}$, $\numhashqmust_1 = 0$, $\gamma_\maxt = \frac{\numhashqmust}{(\maxt-1)\Oraclequeries}$, $\numhashqmust_\maxt = \numhashqmust$, while $\gamma_\IndSect = 0$ and $\numhashqmust_\IndSect = 0$ for $1 <\IndSect < \maxt$.
\end{itemize}

To show this, we use induction on $\maxt$. First, when $C=2$, $W = \gamma_2$. Note that 
\begin{itemize}
\item If $\Oraclequeries \gamma < \numhashqmust$, we can let $\gamma_2 = \gamma$, $\numhashqmust_2 = \numhashqmust$, while $\gamma_1 = 0$ and $\numhashqmust_1 = 0$, in which case $\gamma_2$ is maximized, while all constraints are satisfied; 
\item If $\Oraclequeries \gamma \ge \numhashqmust$, note that for any given $\numhashqmust_2$, $\gamma_2 \le \frac{\numhashqmust_2}{\Oraclequeries} \le \frac{\numhashqmust}{\Oraclequeries}$. When $\numhashqmust \le \Oraclequeries \gamma$, we may let $\gamma_1 = \gamma - \frac{\numhashqmust}{\Oraclequeries}$, $\numhashqmust_1 = 0$, $\gamma_2 = \frac{\numhashqmust}{\Oraclequeries}$, $\numhashqmust_2 = \numhashqmust$, such that all constraints are satisfied, while $\gamma_2$ is maximized.
\end{itemize}
Therefore, our claim is true when $\maxt = 2$. Suppose the claim is true till $\maxt=\inductind$. When $\maxt=\inductind+1$, 
\begin{itemize}
\item If $\inductind \Oraclequeries \gamma < \numhashqmust$, we may  let $\gamma_{\inductind+1} = \gamma$, $\numhashqmust_{\inductind+1} = \numhashqmust$, and let $\gamma_\IndSect = 0$ and $\numhashqmust_\IndSect = 0$ for $1 \le \IndSect \le \inductind$, such that all constraints are satisfied. In this case, $\OPTobj = \inductind \gamma$. Since $\OPTobj \le \inductind \gamma$, we have $\OPTobj^\optimal = \inductind \gamma$. 

\item If $\inductind \Oraclequeries \gamma \ge \numhashqmust $, suppose some $0 \le \gamma' \le \gamma$, $0 \le \numhashqmust' \le \numhashqmust$ are given that also satisfy 
\[
(\gamma-\gamma') \inductind \Oraclequeries \le (\numhashqmust- \numhashqmust') \le (\gamma-\gamma') [(\inductind + 1) \Oraclequeries - 1]
\] 
and
\[
\numhashqmust' \le \gamma' (\inductind \Oraclequeries - 1).
\] 
We can let $\sum_{\IndSect=1}^{\inductind} \gamma_{\IndSect} = \gamma'$ and $\sum_{\IndSect=1}^{\inductind} \numhashqmust_{\IndSect} = \numhashqmust'$. 
\begin{itemize}
\item If $(\inductind-1) \Oraclequeries \gamma' < \numhashqmust'$, based on the induction hypothesis, the maximum value of $\sum_{\IndSect=1}^{\inductind} (\IndSect-1) \gamma_\IndSect$ is $(\inductind-1) \gamma'$, and hence 
\[
\OPTobj \le \inductind \gamma - \gamma'.
\] 
Because 
\[
(\gamma-\gamma') \inductind \Oraclequeries \le (\numhashqmust- \numhashqmust'),
\] 
we have 
\[
\gamma' \ge \gamma - \frac{\numhashqmust}{\inductind \Oraclequeries} + \frac{\numhashqmust'}{\inductind \Oraclequeries}.
\] 
As $(\inductind-1) \Oraclequeries \gamma' < \numhashqmust'$, we have 
\[
\gamma \inductind  - \frac{\numhashqmust}{\Oraclequeries} < \gamma'.
\] 
Therefore, 
\[
\OPTobj \le \frac{\numhashqmust}{\Oraclequeries}.
\] 
\item If $(\inductind-1) \Oraclequeries \gamma' \ge \numhashqmust'$, based on the induction hypothesis, the maximum value of $\sum_{\IndSect=1}^{\inductind} (\IndSect-1) \gamma_\IndSect$ is $\frac{\numhashqmust'}{\Oraclequeries}$, and hence 
\[
\OPTobj \le \frac{\numhashqmust'}{\Oraclequeries} + (\gamma - \gamma') \inductind.
\]
Since $(\gamma - \gamma') \inductind \le \frac{\numhashqmust- \numhashqmust'}{\Oraclequeries}$, we have 
\[
\OPTobj \le \frac{\numhashqmust}{\Oraclequeries}.
\] 
\end{itemize}
Note that $\OPTobj$ achieves  $\frac{\numhashqmust}{\Oraclequeries}$ when $\gamma_1 = \gamma - \frac{\numhashqmust}{\inductind\Oraclequeries}$, $\numhashqmust_1 = 0$, $\gamma_{\inductind+1} = \frac{\numhashqmust}{\inductind\Oraclequeries}$, $\numhashqmust_{\inductind+1} = \numhashqmust$, while $\gamma_\IndSect = 0$ and $\numhashqmust_\IndSect = 0$ for $1 <\IndSect \le \inductind$. 
Therefore, $\OPTobj^\optimal = \frac{\numhashqmust}{\Oraclequeries}$. 
\end{itemize}

Note that actually, in the first case when $\inductind \Oraclequeries \gamma < \numhashqmust$, $\inductind \gamma \le \frac{\numhashqmust}{\Oraclequeries}$, therefore we also have $\OPTobj^\optimal \le \frac{\numhashqmust}{\Oraclequeries}$. Hence,
\[
\OBJ \ge \cfbeta \numhashqmust + \cfgamma \frac{\numhashqmust} {\Oraclequeries} = \frac{\numhashqmust (1-e^{-\Oraclequeries \ratiod})}{\Oraclequeries},
\]
which completes our proof.
\end{IEEEproof}  

Similar to single puzzle case, we may now put things together. Suppose $\BPAlgo$ has an advantage of $\AdverAdv$ in solving the puzzles when receiving $\RecvBits(\BPAlgo)$ bits on average. Based on Theorem ~\ref{thm_compadvmulti}, $\NewAlgo_{\BPAlgo}$ has an advantage of no less than $\AdverAdv-\frac{\adversaries \Oraclequeries}{2^{\Bitsthreshold}}$ while receiving no more than $\RecvBits(\BPAlgo) + \frac{\NumPuzzles \totalsets \setsize \adversaries \Oraclequeries}{2^{\Bitsthreshold}} + \Bitsthreshold \NumPuzzles$  bits on average. Based on Theorem ~\ref{thm_numinformedQmulti}, to achieve an advantage of at least $\AdverAdv-\frac{\adversaries \Oraclequeries}{2^{\Bitsthreshold}}$, any algorithm must make at least $\frac{\NumPuzzles (\AdverAdv - \frac{\Oraclequeries+1}{2^{\Bitsthreshold}})(\totalsets + 1)}{2}$ hash queries. Based on Lemma \ref{thm_multiadvindexavg}, also considering that $\NewAlgo$ needs to receive only $\setsize - \Bitsthreshold + 1$ bits per index set, $\NewAlgo$ receives at least $\nuqidcsavgq(\frac{\NumPuzzles (\AdverAdv - \frac{\Oraclequeries+1}{2^{\Bitsthreshold}}) (\totalsets + 1)}{2}) - \NumPuzzles \totalsets (\Bitsthreshold-1)$ bits on average. Therefore,  
\begin{thm}
Suppose $\adversaries$ adversaries are challenged with $\NumPuzzles$ puzzles. Suppose $\BPAlgo$ solves the puzzle with probability no less than $\AdverAdv$ and let $\RecvBits(\BPAlgo)$ denote the average number of received bits. We have 
\begin{eqnarray}
\RecvBits(\BPAlgo) &\ge& \frac{(1-\shortfrac)\filesize \NumPuzzles (\AdverAdv - \frac{\Oraclequeries+1}{2^{\Bitsthreshold}}) (\totalsets + 1) (1-e^{-\Oraclequeries \setsize / \filesize})}{2 \Oraclequeries} \nonumber \\ 
&& - \NumPuzzles \totalsets (\Bitsthreshold - 1) -  \frac{\NumPuzzles \totalsets \setsize \adversaries \Oraclequeries}{2^{\Bitsthreshold}} - \Bitsthreshold \NumPuzzles \nonumber
\end{eqnarray}
where $\Oraclequeries$, $\Bitsthreshold$, and $\shortfrac$ are constants determined by the puzzle parameters.
\label{thm_multifinal}
\end{thm}

\section{Discussions}
\label{sec:implications}

In this section we discuss the bound and its practical implications. We begin by considering a simple strategy the adversaries may adopt to be compared with the bound.

\subsection{A Simple Adversary Strategy}

We note that there exists a simple strategy the adversaries may adopt to be compared with the bound. In this strategy, when challenged with the puzzles, the adversaries flip a coin and decide whether to attempt to solve the puzzles. They attempt with probability $\AdverAdv$; otherwise they  simply ignore the puzzles. If they decide to solve the puzzles, the adversities select $\frac{\NumPuzzles (\totalsets + 1)}{2 \Oraclequeries}$ members, and let each of them get the entire content. Each of the chosen adversaries makes $\Oraclequeries$ hash queries allowed for them. For each puzzle, the adversaries make hash queries for the index sets one by one until a $\confirm$ is obtained. 

We now analyze the performance of this strategy. We argue that the adversaries can solve the puzzles with probability close to 1 if they decide to attempt, hence their advantage is $\AdverAdv$. Note that to get a $\confirm$ for puzzle according to this strategy, the number of hash queries follows a uniform distribution in $[1,\totalsets]$ and is independent of other puzzles. The total number of hash queries is a random variable with mean $\frac{\NumPuzzles (\totalsets + 1)}{2}$. As the number of puzzles increases, the distribution of this variable approaches a Gaussian distribution centered around the mean with decreasing variance. Therefore, if the adversaries can make $\frac{\NumPuzzles (\totalsets+1)}{2}$ hash queries, the probability that they can solve the puzzles asymptotically approaches 1. Note that this is possible because there are $\frac{\NumPuzzles (\totalsets + 1)}{2 \Oraclequeries}$ selected adversaries, each making  $\Oraclequeries$ queries. According this strategy, the average number of bits downloaded is $\frac{\AdverAdv \filesize \NumPuzzles (\totalsets + 1)}{2 \Oraclequeries}$. 

\subsection{Puzzle Parameter Space}

As can be observed in Theorem~\ref{thm_multifinal}, the dominating factor in the number of bits is roughly  $\frac{\AdverAdv \filesize \NumPuzzles (\totalsets + 1)}{2 \Oraclequeries}$, if the following conditions are satisfied:
\begin{enumerate}
\item $\shortfrac$ is small comparing to 1, 
\item $e^{-\Oraclequeries \setsize / \filesize}$ is small comparing to 1, 
\item $2^{\Bitsthreshold}$ is much larger than $\adversaries \Oraclequeries$, 
\item $2^{\Bitsthreshold}$ is no less than $\setsize \adversaries \Oraclequeries$, 
\item $\Bitsthreshold$ is  much smaller than $\setsize$. 
\vspace{0.02in}
\end{enumerate}
If these conditions are satisfied, the bound approaches the actual number of bits downloaded by the simple adversary strategy above, therefore is tight. 

We show that there are a wide range of values of $\totalsets$, $\setsize$, $\puzzlespercha$, $\filesize$ and $\Oraclequeries$ satisfying these conditions, for which the bound can be applied to provide security guarantees. Note that $\NumPuzzles = \adversaries \puzzlespercha$.  In the following we use an example to illustrate the choice of parameters when $\adversaries \le 10^{6}$; the parameters can be similarly determined for other values of $\adversaries$. Concerning the conditions, 
\begin{itemize}
\item For Condition 1, we note that $\shortfrac$ should be as small as possible, provided that the probability that the number of unique indices in any $\pickseq$ index sets is less than $(1-\shortfrac)\filesize (1-e^{- \pickseq \setsize / \filesize})$ is negligibly small. We find numerically  that when $\adversaries \le 10^{6}$, for $\setsize \ge 10^4$, $\puzzlespercha \totalsets \le 10^{6}$, $\filesize \ge 10^{7}$, if $\shortfrac = 0.1$, this  probability for any $\pickseq$ is below $10^{-12}$. 
\item Condition 2 can be considered as satisfied when $\Oraclequeries \setsize \ge 4 \filesize$, noting that $e^{-4} = 0.018$. 
\item When $\adversaries \le 10^{6}$, $\Bitsthreshold$ can be set to be 60. Condition 3 is satisfied when $ \Oraclequeries \le 10^6$.  Condition 4 is satisfied if $\setsize \le 10^{6}$. Condition 5 is satisfied if $\setsize \ge 10^4$. 
\end{itemize}




The above discussions give the range of the puzzle parameters. Basically, if we let $\shortfrac = 0.1$ and $\Bitsthreshold = 60$, we only require
\comment{
\begin{eqnarray}
&& \filesize \ge 10^7, \label{Cons1}  \nonumber \\
&& 10^{6} \ge \setsize \ge 10^4, \label{Cons2} \nonumber \\
&& \totalsets \puzzlespercha \le 10^6, \label{Cons3} \nonumber \\
&& \Oraclequeries \setsize \ge 4 \filesize, \nonumber \label{Cons4} \\
&& \Oraclequeries \le 10^6, \nonumber \label{Cons5} 
\end{eqnarray}
when $\adversaries \le 10^{6}$. 
}
$\filesize \ge 10^7$, 
$10^{6} \ge \setsize \ge 10^4$,
$10^6 \ge \totalsets \puzzlespercha$, 
$\Oraclequeries \setsize \ge 4 \filesize$, 
$10^6 \ge \Oraclequeries$, 
when $\adversaries \le 10^{6}$. 

\begin{figure}
\begin{center}
\includegraphics[width=2.8in]{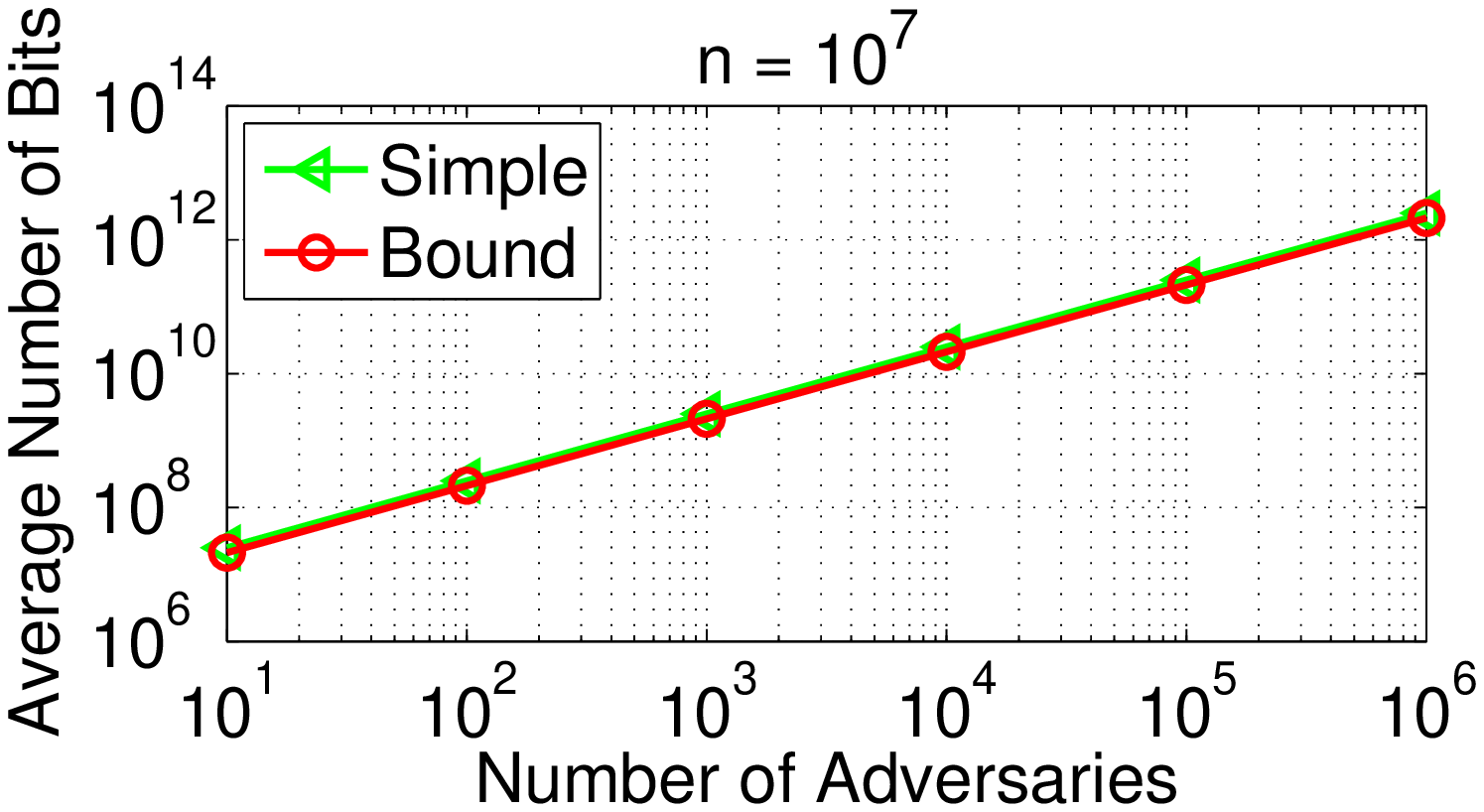} \\
\vspace{-0.1in}
{\small (a)} \\
\includegraphics[width=2.8in]{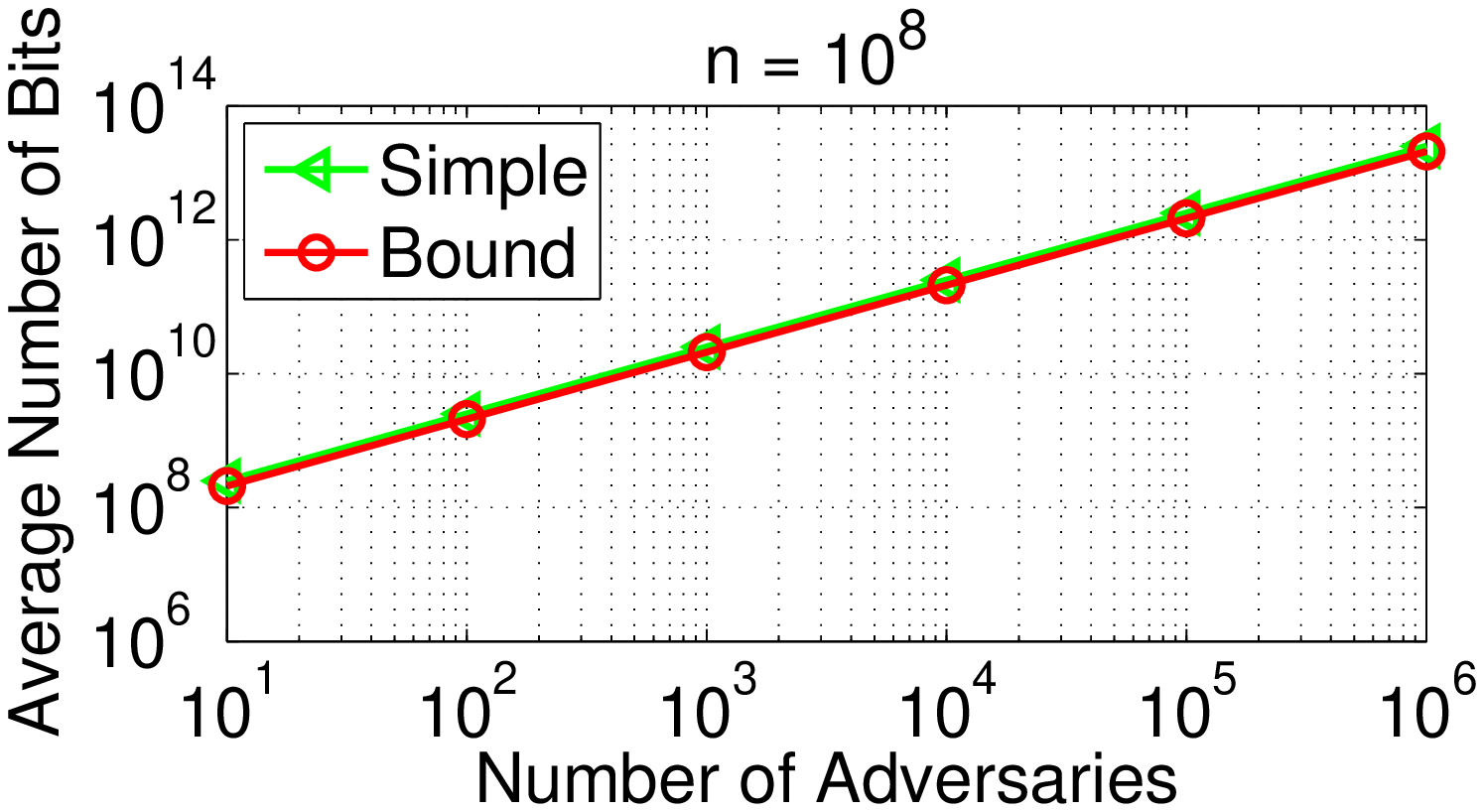} \\
\vspace{-0.1in}
{\small (b)} \\
\end{center}
\vspace{-0.1in}
\caption{Comparison of the average number of bits downloaded by the simple strategy and the bound when $\AdverAdv = 1$, $\setsize = 10^4$, $\Oraclequeries = 4 \filesize / \setsize$ and $\totalsets \puzzlespercha = \Oraclequeries / 2$. (a). $\filesize = 10^7$. (b). $\filesize = 10^8$. }
\vspace{-0.15in}
\label{fig:performance}
\end{figure}

Note that $\setsize$ should be set to its lower bound $10^4$, because a larger value of $\setsize$ results in a heavier load of the verifier. 
$\Oraclequeries$ should be no less than $\totalsets \puzzlespercha$ to ensure an honest prover can solve the puzzles. Figure \ref{fig:performance} shows the average number of bits needed by the simple strategy and the lower bound as a function of the number of adversaries for different content sizes, when $\AdverAdv = 1$, $\Oraclequeries = 4 \filesize / \setsize$ and $\totalsets \puzzlespercha = \Oraclequeries / 2$. We can see that they differ only by a small constant factor. We have tested other parameters satisfying the constraints and the results show similar trends.

\subsection{Puzzle Parameters in Practice}
We also note that the parameter space is not restrictive in practice. Considering the speed of modern communication networks, a reasonable rate to challenge the prover should be once at least every 1MB of data, i.e., when $\filesize$ is at least around $10^7$. There is also no obstacle to set $\setsize$ to be $10^4$ . Concerning $\threshold$ which determines $\Oraclequeries$, note that the puzzles should be solved in a reasonable amount of time to reduce the load of the prover, but the time should also be non-trivial to account for random fluctuations of network latency. Therefore a reasonable value of $\threshold$ should be in the order of several seconds. Given these choices of $\filesize$, $\setsize$, and $\threshold$, $\puzzlespercha$ and $\totalsets$ should satisfy three conditions, according to our earlier discussions: (1) $\setsize \totalsets \le 10^6$, (2) $\setsize \totalsets \puzzlespercha$ should be no less than, for example,  $2\filesize$, and (3) the time to make $\setsize \totalsets$ hash queries should be several seconds but no more than $\threshold$. 

Note that there are two time consuming tasks when making hash queries, which are the hash function call and the generation of the random indices. The choices of hash function and random number generator have been discussed in \cite{ourpuzzlepaper}. Basically, secure hash functions such as SHA-1 can be used as the hash function and block ciphers such as AES can be used to generate the random indices. The optimization of the puzzle implementation is out of the scope of this paper due to the limit of space. We here show the speed of several machines in Emulab \cite{emulab} when executing the SHA-1 hash and the AES encryption in the Openssl library \cite{openssl}, summarized in Table \ref{table:speedres}, when the input to SHA-1 is $10^4$ bits and the AES is 128 bits. If $\filesize = 10^7$, $\setsize = 10^4$ and $\threshold = 3$sec, the results indicate that on modern mainstream machines such as pc3000 and pc2000, (1) it is not possible to make more than $10^6$ hash queries within $\threshold$, (2) it is possible to make enough number of hash queries within $\threshold$ such that $\setsize \totalsets \puzzlespercha \ge 2\filesize$ after optimizations in random index generation, and (3) when $\setsize \totalsets \puzzlespercha \ge 2\filesize$, solving the puzzles will take time in the order of seconds.   

\begin{table}[t]
\begin{centering}
\begin{tabular}{|c|c|c|c|}
\hline 
machine & CPU & SHA1 & AES\\
\hline 
\hline 
pc3000 & 3.0GHz 64-bit & 202165 & 4059157\\
\hline 
pc2000 & 2.0GHz & 71016 & 2605490 \\
\hline  
pc850  & 850MHz & 39151 & 1086667\\
\hline 
pc600  & 600MHz & 29064 & 789624\\
\hline 
\end{tabular}
\par\end{centering}
\caption{SHA-1 and AES function calls executed in one second.}
\vspace{-0.2in}
\label{table:speedres} 
\end{table}

\section{Related Work}
\label{sec:relatedwork}

Using puzzles has been proposed (e.g., in ~\cite{dwork:92,juels:99,memoryboundAbadi,memoryboundDwork,memoryboundJH}) to defend against email spamming or denial of service attacks. In these schemes, the clients are required to spend time to solve puzzles before getting access to the service. The purpose of the bandwidth puzzle is to verify whether the claimed content transactions took place, where the ability to solve the puzzles is tied to the amount contents actually downloaded. As such, the construction of the bandwidth puzzle is different from existing puzzles.

Proofs of data possession (PDP) (e.g.,~\cite{pdpateniese,pdpfilho,pdpscale}) and Proofs of retrievability (POR) (e.g.,~\cite{juelspor,bowerspor,compactpor}) have been proposed to allow a client to verify whether the data has been modified in a remote store. As discussed in \cite{ourpuzzlepaper}, the key differences between PDP/POR schemes and the bandwidth puzzle include the following. First, PDP/POR assumes a single verifier and prover, while the bandwidth puzzle considers one verifier with many potentially colluding provers. Second, the bandwidth puzzle has low computational cost at the verifier, which is desirable in the case when one verifier has to handle many provers, while the existing PDP/POR schemes may incur heavy computational cost at the verifier. The proof techniques for PDP/POR schemes are also different from the techniques used in this paper, because collusion is not considered in existing PDP/POR schemes.

The bandwidth puzzle was first proposed in \cite{ourpuzzlepaper} along with a detailed performance evaluation based on an implemented p2p streaming system and a larger simulated p2p network. The results show that the bandwidth puzzle can effectively improve the performance of the honest users in the presence of colluding adversaries. A bound was also given in \cite{ourpuzzlepaper} on the expected number of puzzles solved when the adversaries can make no more than a certain number of hash queries and download no more than a certain number of content bits. The purpose of this work is to derive a bound on the average number of bits downloaded, when the adversaries can make no more than a certain number of hash queries but can download as many content bits as they wish. Therefore the problem studied in this work is different from that in \cite{ourpuzzlepaper}.  Our analysis show that the new bound is asymptotically tight for all numbers of adversaries, while the bound given in \cite{ourpuzzlepaper} deteriorates quickly as the number of adversaries increases, and the largest number of adversaries used in \cite{ourpuzzlepaper} is 50 when evaluating the bound. As discussed earlier, it is not difficult to find puzzle parameters satisfying the requirements of the new bound. The bound given in \cite{ourpuzzlepaper} requires much more  restrictive choices of parameters. For instance, the suggested values of $\setsize$ and $\totalsets$ are $\frac{1}{4} \filesize^{3/10}$ and $\frac{1}{12} \filesize^{71/100}$, respectively (it was assumed that $\puzzlespercha = 1$ in \cite{ourpuzzlepaper}). A consequence is that the limit on number of bits downloaded by the adversaries must be significantly smaller than $\filesize$ for the bound to give satisfactory results.


\section{Conclusions}
\label{sec:conclusions}

In this paper, we proved a new bound on the performance of the bandwidth puzzle which has been proposed to defend against colluding adversaries in p2p content distribution networks. Our proof is based on reduction, and gives the lower bound of the average number of downloaded bits to achieve a certain advantage by the adversaries. The bound is asymptotically tight in the sense that it is a small fraction away from the average number of bits downloaded when following a simple strategy. The new bound is a significant improvement over the existing bound which was derived under more restrictive conditions and much looser. The improved bound can be used to guide the choice of puzzle parameters to improve the performance of practical systems.


\ignore{

\input{secproof_short}
\input{relwork_short}
\section{Conclusions}
\label{sec:conclusions}

In this paper, we proved a new bound on the performance of the bandwidth puzzle which has been proposed to defend against colluding adversaries in p2p content distribution networks. Our proof is based on reduction, and gives the lower bound of the average number of downloaded bits to achieve a certain advantage by the adversaries. The bound is asymptotically tight in the sense that it is a small fraction away from the average number of bits downloaded when following a simple strategy. The new bound is a significant improvement over the existing bound which was derived under more restrictive conditions and much looser. The improved bound can be used to guide the choice of puzzle parameters to improve the performance of practical systems.

}

{\footnotesize 
\bibliographystyle{abbrv}
\scriptsize
\bibliography{bandpuzzleproof}
}

\end{document}